\documentclass[10pt,twocolumn,reprint,amsmath,amssymb,aps,prl,showpacs,longbibliography,superscriptaddress]{revtex4-1}

\usepackage{color}
\usepackage{amsmath}
\usepackage{amssymb}
\usepackage{graphicx}

\makeatletter

\usepackage{amsfonts}
\usepackage{graphics}

\makeatother

\begin{document}
\title{Discrete time crystals in Bose-Einstein Condensates and symmetry-breaking
edge in a simple two-mode theory}
\author{Jia Wang}
\email{jiawang@swin.edu.au}

\affiliation{Centre for Quantum Technology Theory, Swinburne University of Technology,
Melbourne 3122, Australia}
\author{Krzysztof Sacha}
\affiliation{Instytut Fizyki Teoretycznej, Uniwersytet Jagiello\'{n}ski, ulica
Profesora Stanislawa Lojasiewicza 11, PL-30-348 Krak\'{o}w, Poland}
\author{Peter Hannaford}
\affiliation{Optical Sciences Centre, Swinburne University of Technology, Melbourne
3122, Australia}
\author{Bryan J. Dalton}
\email{bdalton@swin.edu.au}

\affiliation{Centre for Quantum Technology Theory, Swinburne University of Technology,
Melbourne 3122, Australia}
\begin{abstract}
Discrete time crystals (DTCs) refer to a novel many-body steady state
that spontaneously breaks the discrete time-translational symmetry
in a periodically-driven quantum system. Here, we study DTCs in a
Bose-Einstein condensate (BEC) bouncing resonantly on an oscillating
mirror, using a two-mode model derived from a standard quantum field
theory. We investigate the validity of this model and apply it to
study the long-time behavior of our system. A wide variety of initial
states based on two Wannier modes are considered. We find that in
previous studies the investigated phenomena in the evolution time-window
($\lessapprox$2000 driving periods) are actually ``short-time''
transient behavior though DTC formation signaled by
the sub-harmonic responses is still shown if the inter-boson interaction
is strong enough. After a much longer (about 20 times) evolution
time, initial states with no ``long-range'' correlations relax to
a steady state, where time-symmetry breaking
can be unambiguously defined. Quantum revivals also eventually occur.
This long-time behavior can be understood via the many-body Floquet
quasi-eigenenergy spectrum of the two-mode model. A symmetry-breaking
edge for DTC formation appears in the spectrum for strong enough interaction,
where all quasi-eigenstates below the edge are symmetry-breaking
while those above the edge are symmetric. The late-time steady state's
time-translational symmetry depends solely on whether the initial
energy is above or below the symmetry-breaking edge. A phase diagram
showing regions of symmetry-broken and symmetric phases for differing
initial energies and interaction strengths is presented. We
find that according to this two-mode model, the discrete time crystal
survives for times out to at least 250,000 driving periods.
\end{abstract}
\maketitle

\section{Introduction}

Traditionally, spontaneous symmetry breaking (SSB) refers to a situation
where the ground state or a thermal ensemble at a finite temperature
of a many-body system is less symmetrical than its parent Hamiltonian.
Breaking of different symmetries plays a profound role in many aspects
of physics, including (spatial) crystal formation, magnetism, superconductivity
and the origin of particle masses via the Higgs mechanism. However,
spontaneous time-translational symmetry breaking had rarely been considered
until Frank Wilczek proposed the controversial concept of a ``time-crystal''
\citep{Wilczek2012PRL}. Wilczek's original proposal, breaking of
continuous time-translational symmetry in the ground state (or any
thermal equilibrium state), was later rejected by the ``no-go''
theorem of Watanabe and Oshikawa \citep{WatanabePRL2015,Kozin2019PRL}.
Nevertheless, physicists have recently demonstrated that spontaneous
discrete time-translational symmetry breaking (SDTTSB), i.e., discrete
time crystals (DTCs), can exist in out-of-equilibrium systems, such
as Floquet systems that are periodic in time with period $T$ \citep{Sacha2015PRA,Khemani2016PRL,Else2016PRL,Yao2017PRL}.
Experimental evidence of time crystallinity has been reported recently
in a variety of different platforms \citep{Monroe2017Nature,Lukin2017Nature,Barrett2018PRL,Sreejith2018PRL,Barrett2018PRB,Stoof2018PRL,Stoof2019PRA,Stoof2020NJP}.
Discrete time crystals have developed rapidly
recently and have attracted a lot of attention \cite{Else2017PRX,Fazio2017PRB,Sacha2019PRA,Demler2019NJP,Drummond2019NJP,Else2020PRX,Mitra2021PRB,Pizzi2021NP}.
Reviews on the topic of time crystals can be found in Refs. \citep{Sacha2018RPP,Khemani2019arXiv,Yao2020ARCMP,SachaBook2020}.

However, a generic many-body system under periodic driving would normally
keep absorbing energy and approach an infinite temperature state,
a featureless state that cannot support SDTTSB \citep{Marcos2014PRX,Moessner2014PRE,Abanin2015AnnPhys}.
Therefore, the existence of a DTC relies on the prevention (or at
least long-time suppression) of Floquet heating to stabilize the nonequilibrium
quantum state. Indeed, several innovative mechanisms can help to avoid
the heating problem in quantum many-body systems, including many-body
localization (MBL) \citep{Abanin2015PRL,Moessner2015PRL,Abanin2016AnnPhys},
prethermalization \citep{Kuwahara2015AnnPhys,Canovi2016PRE,Yao2019PRR}
and, more recently, many-body quantum scars \citep{Papic2018NP,Serbyn2020PRX,GiudiciPRB2021,Pizz2020iPRB}.

On the other hand, generalizing SSB to time-dependent Floquet systems
where no well-defined ground state or any thermal equilibrium states
exist requires a careful theoretical development \citep{WatanabePRL2015}.
It is now argued that a natural generalization of the equilibrium
notion of SSB can be defined via the long-time steady state \citep{Yao2020ARCMP}.
Following the nomenclature in \cite{Yao2020ARCMP},
steady states in Floquet many-body systems are defined as those with
expectation values of local observables that relax to constants at
stroboscopic times $t=T,\ 2T,\ 3T,...$ In
the case where time-translational symmetry is broken, we extend the
definition to states with constant expectation values of local observables
at $t=sT,\ 2sT,\ 3sT,...$ with $s$ being an integer. In a generic
many-body system, any short-ranged correlated initial state would
usually approach a steady state after some possible transient evolution.
Therefore, in isolated quantum systems out of equilibrium, SSB can
be defined as the situation where the steady state (in the thermalization
limit) is less symmetrical than its parent Hamiltonian. In particular,
a DTC refers to the case where the steady state has a period $sT$
that is a multiple $s$ of the drive period $T$. In the simplest
case, for $s=2$, this amounts to a period-doubling or a subharmonic
response at $0.5\omega\equiv\pi/T$.

A well-known property of Floquet systems is that a time-independent
effective Hamiltonian, namely the Floquet Hamiltonian $H_{F}$, determines
the stroboscopic ($t=T,\ 2T,\ 3T,...$) dynamics \citep{ShirleyPR1965,SambePRA1973,EckardtNJP2015}.
Therefore, the eigenstates and eigenvalues (namely Floquet states
and Floquet energies, respectively) of $H_{F}$ or its equivalent
Floquet evolution operator $U_{F}\equiv\exp(-iH_{F}T)$ also encode
all the necessary information to define SDTTSB. Indeed, Refs. \citep{KhemaniPRB2016,SondhiPRB2016,MoessnerNP2017}
have shown that all eigenstates of $U_{F}$ for a MBL $s=2$ DTC come
in pairs with eigenvalues that have a phase difference $\pi$, which
is associated with the period-doubling of the steady state for systems
starting with physically relevant initial states. In contrast, only
one pair of eigenstates shows the $\pi$-pairing in a many-body quantum
scar system studied in \citep{Pizz2020iPRB}.

Here, we revisit the initial proposal \cite{Sacha2015PRA} that identified
the possibility of realizing
DTCs in a bouncing BEC under periodic driving, where not all but an
extensive set of eigenstates with eigenvalues under a symmetry-breaking
edge come in pairs. This initial work \citep{Sacha2015PRA} and some
of the following studies applies a mean-field approximation \citep{Giergiel2018PRA,Giergiel2020NJP}
or a time-dependent Bogoliubov approximation \citep{KurosNJP2020},
which might artificially preclude or underestimate Floquet heating
in this system since only one or a few modes are included. Mean-field
theories assume that there is no depletion from the condensate mode,
and Bogoliubov theory assumes any depletion is small. As well as allowing
for large depletion and allowing for quantum fluctuations, a multi-mode
treatment is needed to examine the possibility of thermalization,
as this could prevent DTC formation \cite{Khemani2019arXiv,Yao2020ARCMP}.
We recently investigated the case $s=2$ via a fully comprehensive multi-mode quantum treatment
based on a phase-space many-body approach involving the truncated
Wigner approximation (TWA) \citep{WangNJP2021}, which
can include thermalization effects. However, thermalization is found
to be absent in our system, and we find a robust sub-harmonic response
for interactions stronger than some critical value $|g_{c}N|$ and
lasting for a significant period of time, which is a practical criterion
of DTC formation commonly adopted \citep{Khemani2016PRL,Else2016PRL,Yao2017PRL}.
Nevertheless, at present, the TWA is limited in the time regime that
can be computed due to computation resource constraints.

Interestingly, we find for the parameter regime studied
that only two (Wannier) modes were significantly occupied, suggesting
that a many-body theory based on just two dominant modes should work
well in this regime \cite{WangNJP2021}. The two-mode
model allows us to investigate the dynamics at much later evolution
times. In this long-time regime, the system indeed relaxes to a steady
state, where the dynamics are purified and the order parameter for
time-symmetry breaking can be unambiguously determined. References
\citep{Sacha2015PRA,KurosNJP2020} also developed a beyond mean-field
two-mode model. However, the approximation adopted in the derivation
were not clearly justified. Here, we derive the two-mode model from
a standard quantum field theory. We examine the two-mode approximation
in comparison with the multi-mode TWA, and argue that the two-mode
approximation for $s=2$ can be applied to a long evolution time.
We also find that the spectrum of this model Hamiltonian determines
the behavior and symmetry of the long-time steady state. We
note that our two-mode model is not the same as two-mode treatments
in mean-field theory (such as in \cite{Markus2005PRL})
where the condensate wave function is written as a linear combination
of two mode functions. In contrast our two-mode model includes beyond
mean-field effects and allows atoms to macroscopically occupy more
than one mode. We found in our previous multi-mode TWA treatment that
near the critical value of the interaction strength $|g_{c}N|$ (which
corresponds to the onset of DTC formation), the mean-field theory
does not provide a good description of the position probability density
or one-body projector, nor can it account for depletion from the condensate
mode (see Figs 6, 9(b), 9(f) and 11 in Ref. \cite{WangNJP2021}).
The present two-mode model can account for all of these features.

The paper is organized as follows. In Sec. II we introduce the many-body
model for a BEC of bosonic atoms bouncing resonantly on an oscillating
mirror and investigate the validity of the two-mode model for the
case of period-doubling. In Sec. III we present details of our two-mode
model and calculations of the evolution of the system out to very
long times. In Sec. IV we discuss the symmetry breaking in terms of
a symmetry-breaking edge of the two-mode Hamiltonian. Our results
are summarized in Sec. V. Details and some derivations of equations
are given in the Appendices.

\section{Many-body model}

We consider $N$ bosons bouncing vertically on an oscillating mirror
under strong confinement in the transverse directions, which can be
regarded as a one-dimentional (1D) system. The quantum dynamics is
determined by the many-body Schr\"{o}dinger equation $i\hbar\partial t\left|\Theta(t)\right\rangle =\hat{H}\left|\Theta(t)\right\rangle $,
where the Hamiltonian can be written

\begin{equation}
\hat{H}=\int dz\left[\hat{\Psi}(z)^{\dagger}H_{{\rm sp}}\hat{\Psi}(z)+\frac{g}{2}\hat{\Psi}(z)^{\dagger}\hat{\Psi}(z)^{\dagger}\hat{\Psi}(z)\hat{\Psi}(z)\right],\label{eq:Hamiltonian}
\end{equation}
in terms of the field operators $\hat{\Psi}(z)$, $\hat{\Psi}(z)^{\dagger}$
for the annihilation, creation of a bosonic atom of mass $m$ at position
$z$.  Here, the single-particle Hamiltonian is given by $H_{{\rm sp}}=-\partial_{z}^{2}/2+V(z,t)$
using gravitational units for convenience, where the length, time
and energy in gravitational units are $l_{G}=\left(\hbar^{2}/m^{2}g_{E}\right)^{1/3}$,
$t_{G}=\left(\hbar/mg_{E}^{2}\right)^{1/3}$ and $E_{G}=mg_{E}l_{G}$
with $g_{E}$ being the gravitational acceleration. (Throughout this work, quantities shown in the figure are either dimensionless or are given in terms of gravitational units.)
$g=2\omega_{\perp}a_{s}$ is the 1D coupling constant,
where $a_{s}$ is the s-wave scattering length and $\omega_{\perp}$
is the oscillation frequency for the BEC atoms in a transverse trap.
In the coordinate frame moving with the oscillating mirror (which can
be transformed from the laboratory frame via a gauge transformation
\cite{Zakrzewski2002PR,Sacha2015PRA}), the temporally
periodic potential is given by 
\begin{equation}
V(z,t)=z(1-\lambda\cos\omega t),
\end{equation}
with $z\ge0$. Here, $\omega=2\pi/T$ is the driving frequency of
the mirror, and $T$ is the corresponding period. The parameter $\lambda$
determines the driving amplitude. A formal method to solve the single-particle
problem is the Floquet formalism, where the $T$-periodic Floquet
eigenenergies $\epsilon_{\nu}$ and eigenstates $\phi_{\nu}(z,t)=\phi_{\nu}(z,t+T)$
can be defined as 
\begin{equation}
\left[H_{{\rm sp}}-i\partial_{t}\right]\phi_{\nu}(z,t)=\epsilon_{\nu}\phi_{\nu}(z,t).\label{eq:Floquet_sp}
\end{equation}

\begin{figure}
\includegraphics[width=0.98\columnwidth]{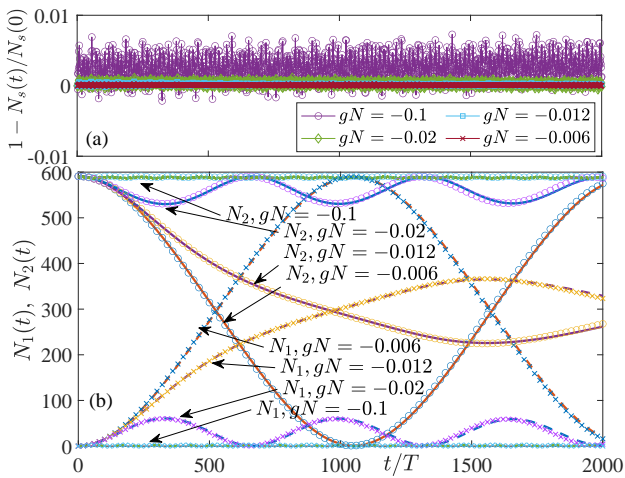}

\caption{(a) TWA calculation of the change in number of atoms in the two Wannier-like
modes $N_{s}=N_{1}+N_{2}$ as a function of time. The initial condition
is realized by preparing a BEC with $N=600$ in a harmonic trap at
initial height $\tilde{h}_{0}=9.82$ and trap frequency $\tilde{\omega}_{0}\approx0.68$.
The change of $N_{s}$ is less than $1\%$ and remains constant within
the fluctuations. (b) Comparison of the occupation
numbers $N_{i}$ of mode $\Phi_{i}$ between the TWA results from
Fig. 10 in Ref. \cite{WangNJP2021} and the two-mode results in this
work. The solid (dashed) line shows the TWA results for $N_{2}$
($N_{1}$), and the circles (crosses) show the two-mode results for
$N_{2}$ ($N_{1}$). The two-mode calculation assumes an initial condition
that $N_{2}=591$ atoms occupy mode $\Phi_{2}(z,0)$. To compare with
the TWA calculation at the critical value $gN=-0.012$, we also need
to use a slightly different value $gN=-0.01185$ in the two-mode model
for the best comparison. \label{fig:TWAresults}}
\end{figure}

The single-particle classical motion under $H_{{\rm sp}}$ is chaotic
for large $\lambda$, and only becomes regular with some suitable
driving parameters and initial conditions. This regular motion can
be recognized by the existence of regular resonance islands in the
classical phase space that are located around periodic orbits with
period $sT$, where $s$ is an integer \cite{Fazio2017PRB,Giergiel2018PRA,Pizzi2021NP}.
Quantum mechanically, such regularity of the classical motion corresponds
to the existence of $s$ special Floquet states $\phi_{\nu}(z,t)$,
where $\nu=1,2,...,s$. Applying a unitary transformation to these
Floquet states, one can construct $s$ Wannier-like states that are
localized both in space and time and with temporal period $sT$ \citep{Giergiel2018PRA}.

In this current work, we focus on the $s=2$ case with $\lambda=0.12$
and $\omega=1.4$ as an example \citep{KurosNJP2020}. The Floquet
states of interest $\phi_{1}(z,t)$ and $\phi_{2}(z,t)$ have quasi-energies
$\epsilon_{1}\approx0.410$ and $\epsilon_{2}\approx1.109$. The two
Wannier-like states are related to the two special Floquet states
via 
\begin{equation}
\begin{array}{l}
\Phi_{1}(z,t)=\frac{1}{\sqrt{2}}\left[\phi_{1}(z,t)+e^{-i\pi t/T}\phi_{2}(z,t)\right]\\
\Phi_{2}(z,t)=\frac{1}{\sqrt{2}}\left[\phi_{1}(z,t)-e^{-i\pi t/T}\phi_{2}(z,t)\right]
\end{array},\label{eq:WannierStates}
\end{equation}
where one can verify that $\Phi_{\nu}(z,t)=\Phi_{\nu}(z,t+2T)$ and
$\Phi_{1}(z,t+T)=\Phi_{2}(z,t)$. These Floquet states and Wannier-like
states have been studied elsewhere \citep{KurosNJP2020,WangNJP2021}
and are plotted in Fig. 1 of Ref. \citep{WangNJP2021}. The key property
we note here is that at $t=0$, $\Phi_{2}(z,t=0)$ is a Gaussian-like
wave-packet. Therefore, if we prepare a weakly interacting BEC confined
in a harmonic trap with suitable initial position $\tilde{h}_{0}$
above the oscillating mirror and trap frequency $\tilde{\omega}_{0}$,
almost all the atoms will initially occupy the mode $\Phi_{2}(z,t=0)$. In our previous TWA calculations in Ref. \cite{WangNJP2021},
we chose $\tilde{h}_{0}=9.82$ and $\tilde{\omega}_{0}=0.68$, and
found that at $t=0$, $N_{2}\approx591$ atoms occupy mode $\Phi_{2}(z,t=0)$
out of a total atom number $N=600$. We used over
50 gravitational modes in our TWA treatment, and listed the other
parameters in the Table 1 of Ref. \cite{WangNJP2021}. This small
difference between $N_{2}$ and $N$ is mainly due to the small mismatch
between $\Phi_{2}(z,t=0)$ and a Gaussian wave-packet. We also find
that the values of $\tilde{h}_{0}$ and $\tilde{\omega}_{0}$ do not
need to be fine-tuned: a finite perturbation does not significantly
reduce $N_{2}$ \citep{KurosNJP2020}.

As pointed out in the Introduction, we have previously studied this
system with finite interaction $g\ne0$ using a multi-mode phase-space
method, namely the truncated Wigner approximation (TWA) \citep{WangNJP2021}.
Interestingly, we find that for $s=2$ during the whole evolution
time $t\le2000T$ only these two Wannier-like modes are occupied.
In Fig. \ref{fig:TWAresults} (a), we show the total occupation of
these two Wannier-like modes $N_{s}(t)=N_{1}(t)+N_{2}(t)$, where
$N_{i}\equiv N_{ii}$ is the occupation number of the mode $\Phi_{i}$.
One can see that the fractional change of $N_{s}(t)$ as a function
of time essentially has just small fluctuations around zero. There
is also no obvious trend of it increasing during the time window investigated.
(In stark contrast, many modes will be occupied if we switch off the
driving for the same initial condition and interaction strength -
see Ref. \citep{WangNJP2021}.) The fluctuations shown in this figure
reflect the stochastic nature of the TWA calculations (and imply the
actual value of $1-N_{s}(t)/N_{s}(0)$ might be smaller than the calculation
uncertainty). In view of only two modes being important in the TWA
calculations for $s=2$, in this work we apply a so-called two-mode
approximation, where we project the Hamiltonian onto the Hilbert sub-space
spanned by Fock states based on the two modes $\Phi_{1}$ and $\Phi_{2}$,
which we name the stable-island Hilbert space. We emphasize here that,
in contrast to fermionic systems that are limited by the Pauli exclusion
principle, our bosonic modes can be occupied by an arbitrary number
of atoms. Therefore, the two-mode model is a generic many-body problem
for large particle number $N$, and the stable-island Hilbert space
has a dimension of $N+1$.

While the two-mode model is described in detail in the next sections,
here in Fig. \ref{fig:TWAresults} (b), we first show the excellent
agreement of the TWA and the two-mode results for different interaction
strengths $gN$. We emphasize that our TWA calculations
\cite{WangNJP2021} show that a DTC forms for interaction strengths
above $g_{c}N=-0.012$ for an initial state prepared in a harmonic
trap with initial position $\tilde{h}_{0}$ and trap frequency $\tilde{\omega}_{0}$.
Quantum fluctuations become significant near this critical interaction
strength $g_{c}N$, and the TWA deviates strongly from the mean-field
results. Nevertheless, the two-mode model is still in excellent agreement
with the TWA in this case. The TWA in principle can include effects
of many (much more than two) modes, and describe the many-body quantum
dynamics exactly in the asymptotic $N\rightarrow\infty$ limit. On
the other hand, although the two-mode model only includes two modes,
it can describe the quantum evolution within the truncated stable-island
Hilbert space exactly for any arbitary $N$. The excellent agreement
between these two models gives us confidence that leakage of atoms
to modes other than $\Phi_{1}$ and $\Phi_{2}$ is negligible during
$t\le2000T$ and most of the atoms seem to be able to remain in the
stable-island Hilbert space for a much longer time. The fact that
most atoms are ``trapped'' in this stable-island Hilbert space for
a long time breaks the ergodicity of the system and prevents Floquet
heating. The underlying physics may be related to a quantum version
of the Kolmogorov--Arnold--Moser (KAM) theory as
pointed out in Ref. \citep{SachaBook2020}, where the existence of
a stable island (as tori in phase space) is robust against weak interactions
\citep{LichtenbergBook1992}. However, non-ergodicity
in KAM theory can be destroyed by an Arnold diffusion after astronomically
long time \citep{SachaBook2020}, implying DTC in
our system has a finite lifetime.

\section{Two-mode approximation}

\begin{figure*}
\includegraphics[width=1.98\columnwidth]{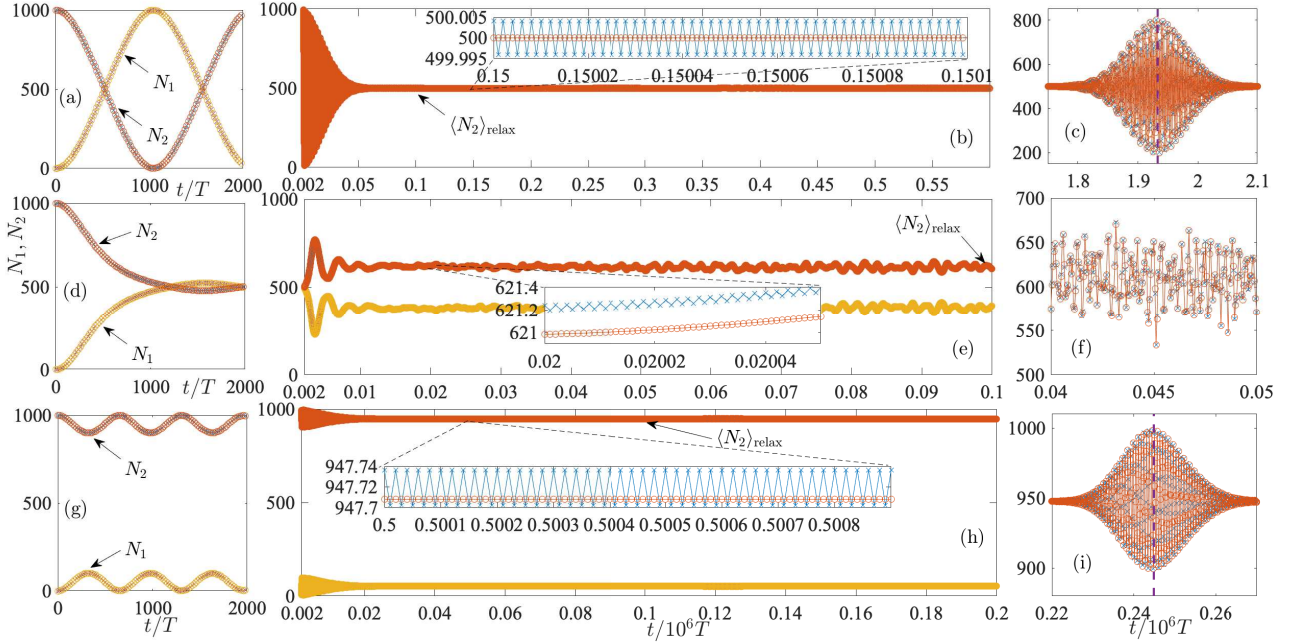}\caption{$N_{1}$ and $N_{2}$ as a function of time for (a-c) weak interaction
$gN=-0.006$, (d-f) near-critical interaction $gN=-0.012$ and (g-i)
strong interaction $gN=-0.02,$ for the initial state $|0,N\rangle$
with total particle number $N=1000$. The crosses (circles) show results
from the MBF approach (HFE approximation). (a) (d) and (g) show the
short-time behavior (at $t=0,20,...,2000T$ for better resolution),
where the HFE approximation works perfectly. (b) (e) and (h) show
the behavior on a much longer time scale, where $N_{2}$ relaxes to
some steady value $\langle N_{2}\rangle_{{\rm relax}}$. For small
interaction $gN=-0.006$, $\langle N_{2}\rangle_{{\rm relax}}=N/2$.
At interaction $gN=-0.012$ near the critical value, stronger fluctuations
occur around the relaxation. At strong interaction $gN=-0.02$, $\langle N_{2}\rangle_{{\rm relax}}>N/2$
without fluctuations. Zoom-ins shown in the insets of (b) (e) and
(h) illustrate minor differences between the MBF and HFE approximation.
(c) (f) and (i) At very long times, a strong quantum revival at around
$t_{{\rm revival}}$ appears for both strong and weak interactions,
but not for interaction near the critical value. Only data for every
$100T$ are shown for clarity. \label{fig:N12_compare}}
\end{figure*}

In the two-mode approximation, we assume the bosonic atom field operator
can be expanded solely by the two Wannier-like modes 
\begin{equation}
\hat{\Psi}(z)=\sum_{i=1,2}\hat{a}_{i}(t)\Phi_{i}(z,t),\ \hat{\Psi}^{\dagger}(z)=\sum_{i=1,2}\hat{a}_{i}^{\dagger}(t)\Phi_{i}^{*}(z,t),\label{eq:TwoModeExpansion}
\end{equation}
where $\hat{a}_{i}(t)$ and $\hat{a}_{i}^{\dagger}(t)$ are the annihilation
and creation operators of a boson in the time-dependent mode $\Phi_{i}(z,t)$.
The time-dependence of the creation operators obeys $a_{i}^{\dagger}(t)=a_{i}^{\dagger}(t+2T)$
and $a_{1}^{\dagger}(t+T)=a_{2}^{\dagger}(t)$, which are determined
by the properties of $\Phi_{i}(z,t)$, and similar rules apply to
the annihilation operators. One can obtain a many-body basis set via
the Fock state \cite{DaltonBook2015}

\begin{equation}
\left|n_{1},n_{2};t\right\rangle =\frac{\left[\hat{a}_{1}^{\dagger}(t)\right]^{n_{1}}}{\sqrt{n_{1}!}}\frac{\left[\hat{a}_{2}^{\dagger}(t)\right]^{n_{2}}}{\sqrt{n_{2}!}}\left|0,0\right\rangle ,
\end{equation}
where $n_{1}$ is the number of atoms in mode $\Phi_{1}$ and
$n_{2}$ is the number of atoms in mode $\Phi_{2}$. Expanding the many-body
state vector in this basis set gives
\begin{equation}
\left|\Theta(t)\right\rangle =\exp(-i\mu t)\sum_{n=0}^{N}b_{n}(t)\left|n,N-n;t\right\rangle ,
\end{equation}
where the total number of bosonic atoms $N$ is a good quantum number
and $\mu$ is a suitable frequency (chosen below). These Fock states
also satisfy $\left|n_{1},n_{2};t+2T\right\rangle =\left|n_{1},n_{2};t\right\rangle $
and $\left|n_{1},n_{2};t+T\right\rangle =\left|n_{2},n_{1};t\right\rangle $.
We note that, while both the creation/annihilation operators and the
Fock-state basis are time-dependent, the matrix elements of any direct
product of creation and annihilation operators at the same time are
time-independent, e.g., $\left\langle n_{1}+1,n_{2}-1;t\right|\hat{a}_{1}^{\dagger}(t)\hat{a}_{2}(t)\left|n_{1},n_{2};t\right\rangle =\sqrt{\left(n_{1}+1\right)n_{2}}$.

The time-evolution of the expansion coefficients $b_{n}(t)$ is determined
by the many-body Schr{ö}dinger equation, in a vector form 
\begin{equation}
\left[\underline{\widetilde{H}}(t)-i\hbar\partial_{t}\right]\vec{b}(t)=0,\label{eq:EvolutionEq}
\end{equation}
where $\underline{\widetilde{H}}(t)$ is a time-dependent matrix with
matrix elements $\widetilde{H}_{mn}(t)=\left\langle m,N-m;t\right|\hat{\mathcal{H}}(t)\left|n,N-n;t\right\rangle $.
Here, the effective Hamiltonian operator is given by 
\begin{equation}
\hat{\mathcal{H}}(t)=J\left(\hat{a}_{1}^{\dagger}\hat{a}_{2}+{\rm h.c.}\right)+\frac{1}{2}g\sum_{ijkl=1}^{2}U_{ijkl}(t)\hat{a}_{i}^{\dagger}\hat{a}_{j}^{\dagger}\hat{a}_{k}\hat{a}_{l},\label{eq:H2tmode}
\end{equation}
where $J=(\mathcal{\epsilon}_{1}-\mathcal{\epsilon}_{2}+\hbar\omega/2)/2$
and $U_{ijkl}(t)=\int dz\Phi_{i}^{*}(z,t)\Phi_{j}^{*}(z,t)\Phi_{k}(z,t)\Phi_{l}(z,t)$.
The phase factor $\mu$ occurring in the quantum state $\left|\Theta(t)\right\rangle $
is given by $\mu=N\epsilon/\hbar$, where $\epsilon=(\mathcal{\epsilon}_{1}+\mathcal{\epsilon}_{2}-\hbar\omega/2)/2.$
The derivation is given in Appendix \ref{subsec:Derivation1}. With
an understanding that the creation and annihilation operators will
always act on the Fock basis and introducing time-independent matrix
elements, we hereafter leave implicit the time-dependence of $\hat{a}_{i}$
and $\hat{a}_{i}^{\dagger}$. The only explicit time-dependence remains
in the parameter $U_{ijkl}(t)$, which is periodic with period $\widetilde{T}=2T$
and $\widetilde{\omega}=\omega/2$. The effective Hamiltonian matrix
$\underline{\widetilde{H}}(t)$ also shares the same periodicity,
implying that Eq. (\ref{eq:EvolutionEq}) can be formally solved using
a Floquet approach. A many-body Floquet state and Floquet energy can
be defined as 
\begin{equation}
\left[\underline{\widetilde{H}}(t)-i\hbar\partial_{t}\right]\vec{\mathcal{F}}_{\nu}(t)=\tilde{\mathcal{E}}_{\nu}\vec{\mathcal{F}}_{\nu}(t),\label{eq:Full2mode}
\end{equation}
and the time evolution can be obtained via $\vec{b}(t)=\sum_{\nu}\mathcal{C}_{\nu}\vec{\mathcal{F}}_{\nu}(t)e^{-i\tilde{\mathcal{E}}_{\nu}t}$
and $\mathcal{C}_{\nu}=\vec{\mathcal{F}}_{\nu}^{\dagger}(t=0)\vec{b}(0)$
(see Appendix \ref{sec:MBF}). We emphasize here that this approach,
namely the many-body Floquet (MBF) approach, is a full and exact many-body
quantum calculation as long as only two modes are occupied and depletion
to other modes is negligible.

Another time scale in the Hamiltonian is given by $t_{{\rm tunneling}}=1/J$
which describes the tunneling of a single particle between the two
modes. $t_{{\rm tunneling}}$ typically is in the order of $500T$,
much longer than $2T$, the period of $U_{ijkl}(t)$. An approximation
to simplify and understand this problem is therefore to take a high-frequency
expansion (HFE) of the Hamiltonian matrix $\underline{\widetilde{H}}(t)=\sum_{\delta}\underline{h}_{\delta}e^{-i\delta\tilde{\omega}t}$,
where $\underline{h}_{\delta}=\int_{0}^{\tilde{T}}dt\underline{\widetilde{H}}(t)e^{i\delta\tilde{\omega}t}/\tilde{T}$,
and keep only the lowest order $\underline{h}_{0}$ - which is a time-independent
effective Hamiltonian matrix. The corresponding Hamiltonian operator
$\hat{h}_{0}$ of $\underline{h}_{0}$ agrees with the one given in
Ref. \citep{Sacha2015PRA}, which can be expressed in the same form
as Eq. (\ref{eq:H2tmode}) with $U_{ijkl}(t)$ replaced by the $2T$-average
value $\bar{U}_{ijkl}=\int_{0}^{2T}U_{ijkl}(t)dt/2T$. However, we
note that the time-independent Hamiltonian in Ref. \citep{Sacha2015PRA}
is derived by expanding and truncating the Floquet Hamiltonian and
field operators in an extended space-time Hilbert space. This expansion
and truncation is thus an approximation that is effectively equivalent
to our HFE approximation. Under the HFE approximation, the time evolution
can be approximately given by $\vec{b}(t)=\sum_{\nu}c_{\nu}\vec{f}_{\nu}e^{-i\mathcal{E}_{\nu}t}$
and the initial condition $c_{\nu}=\vec{f}_{\nu}^{\dagger}\vec{b}(0)$.
Here, $\mathcal{E}_{\nu}$ and $\vec{f}_{\nu}$ are eigenenergies
and eigenstates of $\underline{h}_{0}$: $\underline{h}_{0}\vec{f}_{\nu}=\mathcal{E}_{\nu}\vec{f}_{\nu}$.

Discussions on SDTTSB usually refer to initial states without extensive
many-body quantum entanglement, which are physically accessible by
simple preparation schemes. As mentioned before, by preparing a BEC
in a harmonic trap with suitable potential height and width, the initial
condition can be well approximated by a product state $|0,N\rangle$.
If we turn off the interaction in the preparation stage, the particles
can tunnel from $\Phi_{2}$ to $\Phi_{1}$ with a tunneling rate $J$,
and an additional relative phase can also be induced by some phase-printing
scheme. Therefore, in principle, we can prepare a BEC initial product
state, which in first quantization is 
\begin{equation}
|\{\theta,\varphi\}\rangle=|\Lambda_{2}\rangle_{1}|\Lambda_{2}\rangle_{2}...|\Lambda_{2}\rangle_{N},\label{eq:theta_state}
\end{equation}
where all atoms occupy a single mode $\Lambda_{2}=\sin\theta e^{i\varphi}\Phi_{1}+\cos\theta\Phi_{2}$.
Without loss of generality, we choose $\theta\in[0,\pi)$ and $\varphi\in[0,\pi)$.

To investigate the dynamical evolution, we study the observables $N_{ij}(t)=\langle\Theta(t)|\hat{a}_{i}^{\dagger}\hat{a}_{j}|\Theta(t)\rangle$,
which play a similar role to the one-body reduced density matrix elements.
For simplification of notation, we define $N_{i}(t)\equiv N_{ii}(t)$.
Figure \ref{fig:N12_compare} shows a comparison between the MBF approach
and HFE approximation for $gN=-0.006,\ -0.012$ and $-0.02$. The
initial state is chosen as $|0,N\rangle$ wih $N=1000$. Figure \ref{fig:N12_compare}
(a) (d) and (g) show perfect agreement between the HFE and MBF at
relatively short time scales (0 to 2000$T$). At longer time scales
$t>t_{{\rm relax}}$, Fig. \ref{fig:N12_compare} (b) (e) and (h)
show that $N_{2}(t)$ relaxes to a constant $\langle N_{2}\rangle_{{\rm relax}}$
for the HFE. However, for an interaction strength near the critical
value $gN=-0.012$, $N_{2}(t)$ has a relatively larger fluctuation.
The MBF results here also show excellent agreement with the HFE, except
for an almost negligible oscillation around the $\langle N_{2}\rangle_{{\rm relax}}$
that can only be seen in the insets for a large zoom-in scale. This
small oscillation can be understood as micro-motion originating from
the time-dependence of the many-body Floquet states in the MBF approach.
After an even longer evolution time, Fig. \ref{fig:N12_compare} (c)
and (i) show a strong quantum revival for both strong and weak interaction,
but no revival can be seen for the critical case, Fig. \ref{fig:N12_compare}
(f). We also note that for $gN=-0.02$, the quantum revival seems
to be almost perfect, i.e. $N_{2}(t_{{\rm revival}})\approx N$, in
contrast to the case for $gN=-0.006$ where $N_{2}(t_{{\rm revival}})<N$.
Visible differences between the MBF and HFE show up in the quantum
revival regime. Nevertheless, the HFE still accurately predicts the
revival time $t_{{\rm revival}}$ and the overall revival profile.
As one can see, the HFE is an excellent approximation. The HFE will
also provide an insight into the symmetry breaking and give an analytical
explanation of the time-evolution behavior, as shown in the next section.

While the quantum revival is interesting and indicates there is a
relatively short transient period of deviation from an almost perfect
temporal periodicity, the revival regime is much shorter than the
steady-state regime. We discuss the quantum revivals with more detail
in Appendix \ref{sec:revival} and focus on the steady-state regime
here. We emphasize that this relaxation value of $\langle N_{2}\rangle_{{\rm relax}}$
reflects the time-translational symmetry of the steady state. For
example, the quantum correlation function (QCF) $P\left(z,z^{\#},t\right)={\rm Tr}\left[\hat{\Psi}^{\dagger}\left(z^{\#}\right)\hat{\Psi}(z)\left|\Theta(t)\right\rangle \left\langle \Theta(t)\right|\right]$
is given by 
\begin{equation}
P(z,z^{\#},t)=\sum_{i,j=1}^{2}N_{ji}(t)\Phi_{i}(z,t)\Phi_{j}^{*}(z^{\#},t),
\end{equation}
where $z$ and $z^{\#}$ refer to two different position coordinates.
The position probability density at stroboscopic times $t=kT$ with
$k=1,2,3...$ can be well approximated by $F(z,kT)=P(z,z,kT)\approx N_{1}(kT)\left|\Phi_{1}(z,kT)\right|^{2}+N_{2}(kT)\left|\Phi_{2}(z,kT)\right|^{2}$,
where the cross-terms can be neglected since $\Phi_{1}$ and $\Phi_{2}$
are well separated localized wave-packets at $t=kT$ with $k$ being
integer. We recall that $\Phi_{1}(z,kT)=\Phi_{2}(z,kT+T)$. Therefore,
if $\langle N_{2}(t)\rangle\rightarrow\langle N_{2}\rangle_{{\rm relax}}=N/2$
as in Fig. \ref{fig:N12_compare} (b), $F(z,kT)=F(z,kT+T)$ has the
same period as the Hamiltonian, which represents a symmetry-unbroken
state. On the other hand, if $\langle N_{2}(t)\rangle\rightarrow\langle N_{2}\rangle_{{\rm relax}}\ne N/2$
as in Fig. \ref{fig:N12_compare} (h), $F(z,kT)\ne F(z,kT+T)$ and
$F(z,kT)=F(z,kT+2T)$, indicating that the system's state breaks the
discrete time-translational symmetry, i.e., a time crystal is present.
Figures 2 (h) and (i) predict that the time crystal
survives for times out to at least 250,000 driving periods.

We can also define a one-body projection operator (OBP) $\hat{M}_{G}$
onto the Gaussian-like localized wave-packet $\Phi_{2}(z,0)$ as an
observable \citep{WangNJP2021}. In the first quantization, the OBP
is given by $\hat{M}_{G}=\sum_{i=1}^{N}(|\chi\rangle\langle\chi|)_{i}$,
where $\langle z|\chi\rangle=\Phi_{2}(z,0)$ and $i$ lists individual
particles as $i=1,2,...,N$. The expectation value of the OBP can
be related to the QCF $\hat{\langle\Psi}\left(z^{\#}\right)^{\dagger}\hat{\Psi}(z)\rangle$
via $M_{G}(t)=\iint\mathrm{d}z\mathrm{~d}z^{\#}\Phi_{2}\left(z^{\#},0\right)\hat{\langle\Psi}\left(z^{\#}\right)^{\dagger}\hat{\Psi}(z)\rangle\Phi_{\mathrm{2}}^{*}(z,0)$.
At stroboscopic times $t=kT$ with $k$ being integer, the OBP can
be simplified as

\begin{equation}
M_{G}(kT)=\begin{cases}
N_{1}(kT)=N-N_{2}(kT) & k=1,3,5,...\\
N_{2}(kT) & k=0,2,4,...
\end{cases}
\end{equation}
The physical picture is clear: $M_{G}(kT)$ measures how many atoms
occupy the Gaussian-like wave-packet $\Phi_{2}(z,0)$ at stroboscopic
times. When $k$ is even {[}odd{]}, this Gaussian-like wave-packet
coincides with $\Phi_{2}(z,kT)$ {[}$\Phi_{1}(z,kT)${]}. Therefore,
via the investigation of the behavior of $N_{2}(kT)$, we can obtain
the evolution of the observable $M_{G}(kT)$, which reveals the temporal
symmetry of the steady state. If $\langle N_{2}\rangle_{{\rm relax}}=N/2$,
then $M_{G}(kT)\rightarrow N/2$ implies a $T$-periodicity. On the
contrary, if $\langle N_{2}\rangle_{{\rm relax}}\ne N/2$, $M_{G}(kT)$
has $2T$-periodicity, indicating SDTTSB. One can define a frequency
response observable via the Fourier transformation

\begin{equation}
m_{G}(f)=\frac{1}{K}\sum_{k=k_{0}}^{k_{0}+K-1}e^{-ifkT}\frac{M_{G}(kT)}{N},
\end{equation}
where we usually choose $K=2048$ and $k_{0}=10^{4}$ so that $k_{0}T$
is in the steady-state regime. In particular, the sub-harmonic response
is given by

\begin{equation}
|m_{G}(\omega/2)|\rightarrow\frac{|N-2\langle N_{2}\rangle_{{\rm relax}}|}{2N}=\frac{P_{G}}{2},
\end{equation}
where $P_{G}=|\langle N_{2}\rangle_{{\rm relax}}-\langle N_{1}\rangle_{{\rm relax}}|/N$
is the population imbalance. $|m_{G}(\omega/2)|=0$ (or equivalently
$P_{G}=0$) indicates the symmetry-unbroken phase, and hence $|m_{G}(\omega/2)|$
can be regarded as an order parameter.

\section{Symmetry breaking}

\begin{figure}
\includegraphics[width=0.98\columnwidth]{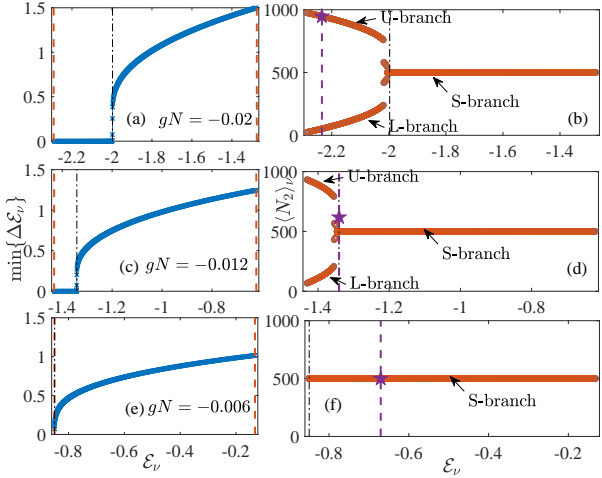}\caption{Symmetry-breaking edge of the two-mode Hamiltonian with $N=1000$
as a function of eigenenergy $\mathcal{E}_{\nu}$. (a) (c) and (e)
show the minimum of the gap ${\rm min}\{\Delta\mathcal{E}_{\nu}\}$
between adjacent eigenenergies for $gN=-0.02,\ -0.012,\ $ and $-0.006$,
respectively. The symmetry-breaking edge indicated by the black dash-doted
vertical line and the range of eigenenergies are given by $E_{{\rm min}}$
and $E_{{\rm max}}$ indicated by the red dashed vertical lines on
the left and right. ${\rm min}\{\Delta\mathcal{E}_{\nu}\}$ is essentially
zero for a near-degenerate pair as shown in (a) and (c) for eigenenergies
below the edge. In (e), the edge is about the same as the minimum
of the spectrum $E_{{\rm min}}$ and hence no near-degenerate pair
exists. (b) (d) and (f) show $\langle N_{2}\rangle_{\nu}$ in Eq.
(\ref{eq:N2nu}) as a function of $\mathcal{E}_{\nu}$, where the
onset of bifurcation occurs at the edge indicated by the black vertical
dash-dotted line. We name the eigenstates with no bifurcation of $\langle N_{2}\rangle_{\nu}=N/2$
the S-branch and $\langle N_{2}\rangle_{\nu}>N/2$ ($\langle N_{2}\rangle_{\nu}<N/2$)
in the bifurcation regime the U-branch (L-branch). The purple dashed
line shows the initial energy $E_{{\rm ini}}$ for the initial state
$|0,N\rangle$, and the purple pentagram shows the corresponding relaxation
value $\langle N_{2}\rangle_{{\rm relax}}$. \label{fig:Edge}}
\end{figure}

\begin{figure}
\includegraphics[width=0.98\linewidth]{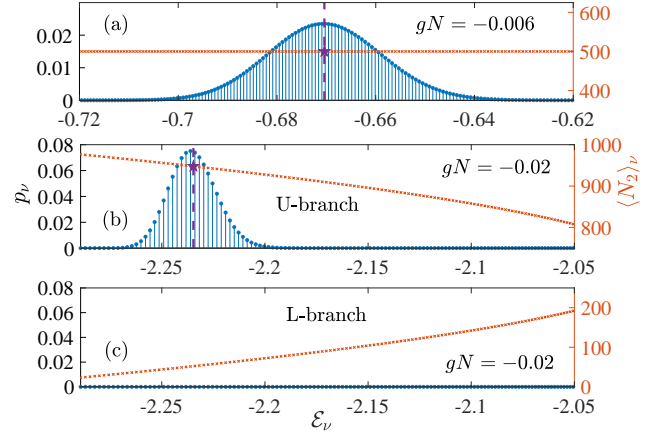}\caption{Projection $p_{\nu}=|c_{\nu}|^{2}$ of initial state $|0,N\rangle$
with $N=1000$ as a function of $\mathcal{E}_{\nu}$ (blue circles)
and the corresponding $\langle N_{2}\rangle_{\nu}$ (red crosses).
(a) shows the projection for the case $gN=-0.006$, which is only
significant around the initial energy $E_{{\rm ini}}$ indicated by
the purple dashed vertical line. The diagonal ensemble prediction
$\sum_{\nu}p_{\nu}\langle N_{2}\rangle_{\nu}=N/2$ is indicated by
the purple pentagram. (b) and (c) show the projection to the U- and
L-branch, respectively, for $gN=-0.02$. The projection is only significant
for the U-branch around the initial energy $E_{{\rm ini}}$. The diagonal
ensemble prediction is indicated by the purple pentagram. \textcolor{black}{\label{fig:Projection}}}
\end{figure}

Under the HFE approximation, the effective time-independent Hamiltonian
satisfies the $\mathbb{Z}_{2}$ symmetry determined by the operator
$\hat{P}_{12}=i^{N}\exp\left[i\pi(\hat{a}_{1}^{\dagger}\hat{a}_{2}+\hat{a}_{2}^{\dagger}\hat{a}_{1})/2\right]$
which interchanges the mode indices $1\leftrightarrow2$. We find
that $\hat{P}_{12}\hat{a}_{1}\hat{P}_{12}^{-1}=i\hat{a}_{2}$ and
$\hat{P}_{12}\hat{a}_{2}\hat{P}_{12}^{-1}=i\hat{a}_{1}$. The corresponding
effective Hamiltonian operator can be rewritten as 
\begin{equation}
\begin{alignedat}{1}\hat{h}_{0} & =J\left(\hat{a}_{1}^{\dagger}\hat{a}_{2}+{\rm h.c.}\right)+gu_{T}\left(\hat{a}_{1}^{\dagger}\hat{N}\hat{a}_{2}+{\rm h.c.}\right)\\
 & +\frac{1}{2}g\left[u_{I}\sum_{i=1}^{2}\hat{N}_{i}(\hat{N}_{i}-1)+4u_{N}\hat{N}_{1}\hat{N}_{2}\right]\\
 & +\frac{1}{2}gu_{P}\left(\hat{a}_{1}^{\dagger}\hat{a}_{1}^{\dagger}\hat{a}_{2}\hat{a}_{2}+{\rm h.c.}\right),
\end{alignedat}
\label{eq:h0}
\end{equation}
where $u_{T}=\bar{U}_{1112}$, $u_{I}=\bar{U}_{1111}$, $u_{N}=\bar{U}_{1212}$
and $u_{P}=\bar{U}_{1122}$ are the only four distinctive values of
$\bar{U}_{ijkl}$ constrained by the $\mathbb{Z}_{2}$ symmetry (see
Appendix \ref{subsec:Derivation2}). The effective Hamiltonian $\hat{h}_{0}$
is invariant under the symmetry operator $\hat{P}_{12}.$ Here, the
number operators are given by $\hat{N}_{1}=\hat{a}_{1}^{\dagger}\hat{a}_{1}$,
$\hat{N}_{2}=\hat{a}_{2}^{\dagger}\hat{a}_{2}$ and $\hat{N}=\hat{N}_{1}+\hat{N}_{2}$.
One might notice, under a mean-field approximation,
that this effective Hamiltonian can be applied to investigate the
bosonic self-trapping phenomenon, for example, in a BEC in a double-well
potential \cite{Markus2005PRL}. The connection of DTC formation and
self-trapping was recognized in Ref. \cite{Sacha2015PRA}. Here, we
go beyond the mean-field approximation, and numerically investigate
the whole eigen spectrum exactly. In our numerical investigation
here, we have $J\approx3.580\times10^{-4}$, $u_{I}\approx0.2237$,
$u_{N}\approx0.0519$, $u_{T}\approx-1.9\times10^{-4}$ and $u_{P}\approx-4.3\times10^{-6}$.
Since the spectrum of $\hat{h}_{0}$ is bounded from both below and
above, the maximum and minimum eigenenergy of the $\hat{h}_{0}$ can
be obtained via the mean-field approach in the large $N$ limit. Neglecting
the term associated with $u_{P}$ (which is much smaller than $u_{T}$,
$u_{I}$ and $u_{N}$), this approach leads to 
\begin{equation}
E_{{\rm max}}\approx E_{{\rm shift}}+|\tilde{J}|N/2
\end{equation}
and 
\begin{equation}
E_{{\rm min}}\approx\begin{cases}
E_{{\rm shift}}-|\tilde{J}|N/2, & |gN|\le|g_{b}N|\\
\frac{1}{2}gN^{2}u_{I}+\frac{\tilde{J}^{2}N}{gN(u_{I}-2u_{N})}, & |gN|>|g_{b}N|
\end{cases},
\end{equation}
where $g_{b}N\equiv-2J/(u_{I}-2u_{N}+2u_{T})$. (We use the conditions
$J>0$ and $gN<0$ here. We also focus on the case $J+gNu_{T}>0$
since $|gN|$ is small. See Appendix \ref{sec:mf} for details.) Here,
$E_{{\rm shift}}=gN\left(u_{I}N/2-u_{I}+u_{N}N\right)/2$ and its
physical meaning will be clear below.

For a given total number $N$ (since $\hat{N}$ commutes with $\hat{h}_{0}$),
the Hamiltonian can be mapped to the Lipkin--Meshkov--Glick (LMG)
model (see Appendix \ref{subsec:Derivation3}) as \citep{Ribeiro2008PRE,Fabrizio2012PRB,KurosNJP2020}
\begin{equation}
\hat{h}_{0}\approx\tilde{J}\left(-\hat{S}_{x}+\frac{\gamma}{N}\hat{S}_{z}^{2}\right)+E_{{\rm shift}},\label{eq:LMKH}
\end{equation}
with $\hat{S}_{x}=(\hat{a}_{1}^{\dagger}\hat{a}_{2}+\hat{a}_{2}^{\dagger}\hat{a}_{1})/2$
and $\hat{S}_{z}=(\hat{a}_{2}^{\dagger}\hat{a}_{2}-\hat{a}_{1}^{\dagger}\hat{a}_{1})/2,$
which are bosonic spin operators. The LMG Hamiltonian
can be applied to describe spin-systems with infinite-range interaction.
However, we emphasize here that our system consists of bosonic atoms
with zero-range contact interactions. When the two-mode and HFE approximations
are applicable, the time-independent LMG Hamiltonian in Eq. (\ref{eq:LMKH})
becomes an excellent approximation for the Floquet Hamiltonian of
the underlying physical system, and determines the dynamical properties
of the original system. The LMG Hamiltonian is invariant under the
symmetry operator $\hat{P}_{12}.$ The parameters are given by $\tilde{J}=-2\left[J+gu_{T}(N-1)\right]$,
$\gamma=gN\left(u_{I}-2u_{N}\right)/\tilde{J}$. In the case of large
$N$ and finite $\gamma\propto gN$, a symmetry-broken edge exists
for the LMG Hamiltonian \cite{Ribeiro2008PRE,Fabrizio2012PRB,Fazio2017PRB,KurosNJP2020}:
\begin{equation}
\mathcal{E}_{{\rm edge}}\approx-|\tilde{J}|N/2+E_{{\rm shift}}.\label{eq:Eedge}
\end{equation}
For $\mathcal{E}_{\nu}>\mathcal{E}_{{\rm edge}}$ the eigenenergies
are non-degenerate whilst those for $\mathcal{E}_{\nu}<\mathcal{E}_{{\rm edge}}$
are essentially two-fold degenerate (see Fig. \ref{fig:Edge}). From
the expression for $E_{{\rm min}}$ and $\mathcal{E}_{{\rm edge}}$,
one can observe that $E_{{\rm min}}=\mathcal{E}_{{\rm edge}}$ for
weak interaction $|gN|<|g_{b}N|$, where $g_{b}N$ is defined above
and Eq. (\ref{eq:gbN}). For strong interaction $|gN|>|g_{b}N|$,
$E_{{\rm min}}<\mathcal{E}_{{\rm edge}}<E_{{\rm max}}$ and the eigenpairs
below and above the edge have distinct features. We emphasise that,
according to group representation theory, since $\mathbb{Z}_{2}$
is the symmetry group, the eigenstates for $\hat{h}_{0}$ would in
general be non-degenerate and either even or odd under the symmetry
operator, i.e., $\hat{P}_{12}|\nu,\pm\rangle=\pm|\nu,\pm\rangle$.
However, for $|gN|>|g_{b}N|$, eigenstates that satisfy $\mathcal{E}_{\nu}<\mathcal{E}_{{\rm edge}}$
form near-degenerate pairs $|\nu,+\rangle$ and $|\nu,-\rangle$ of
opposite symmetry, with an energy gap $|\mathcal{E}_{\nu}^{(+)}-\mathcal{E}_{\nu}^{(-)}|$
that is exponentially small in $N$. This energy gap is negligible
for a large and finite $N$ in practice, and becomes exactly zero
in the infinite $N$ (thermodynamic) limit. Therefore, we can choose
a pair of symmetry-broken states that satisfy $\hat{P}_{12}\left|\nu\right\rangle _{U}=\left|\nu\right\rangle _{L}$
and $\hat{P}_{12}\left|\nu\right\rangle _{L}=\left|\nu\right\rangle _{U}$,
which are given by $|\nu\rangle_{U}=(|\nu,+\rangle+|\nu,-\rangle)/\sqrt{2}$
and $|\nu\rangle_{L}=$$(|\nu,+\rangle-|\nu,-\rangle)/\sqrt{2}$.
$|\nu\rangle_{U}$ and $\left|\nu\right\rangle _{L}$ have the same
expectation energies $\mathcal{E}_{\nu}^{(U)}=\mathcal{E}_{\nu}^{(L)}$
and serve as a degenerate pair of symmetry-broken eigenstates. In
contrast, any eigenstates with eigenenergies $\mathcal{E}_{\nu}>\mathcal{E}_{{\rm edge}}$
show no near-degeneracy and these states $\left|\nu\right\rangle _{S}$
are symmetry-unbroken. Figures \ref{fig:Edge} (a), (c) and (e) show
the symmetry-breaking edge for $gN=-0.02$, $-0.012$ and $-0.006$,
respectively. The vertical axis shows the minimum of adjacent gaps:
$\min\left\{ \Delta\mathcal{E}_{\nu}\right\} =\min(\mathcal{E_{\nu}}-\mathcal{\mathcal{E}}_{\nu-1},\mathcal{E}_{\nu+1}-\mathcal{\mathcal{E}}_{\nu})$,
which is exactly zero if and only if there is a double-degeneracy.
One can see that this quantity indeed becomes essentially zero at
$\mathcal{E}_{{\rm edge}}$, which is denoted by the black vertical
dash-dotted lines. The red dashed line on the right (left) indicates
$\mathcal{E}_{{\rm min}}$ ($\mathcal{E}_{{\rm max}}$). For $|gN|\le|g_{b}N|\approx0.006$,
only a symmetry-unbroken phase exists, since $\mathcal{E}_{{\rm edge}}\approx\mathcal{E}_{{\rm min}}$,
and no near-degenerate pairs of energy levels occur.

\begin{figure}
\includegraphics[width=0.98\columnwidth]{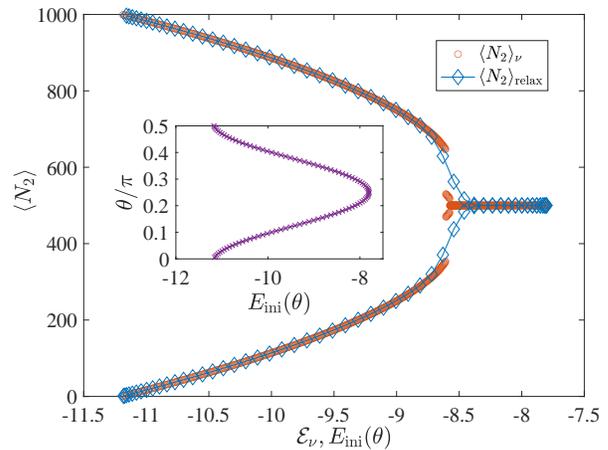}

\caption{$\langle N_{2}\rangle_{{\rm relax}}$ (blue diamonds) for
different initial states as a function of the initial energy $E_{{\rm ini}}(\theta)\equiv E_{{\rm ini}}(\theta,\varphi=0)$
in Eq. (\ref{eq:Eini}) for $gN=-0.1$. Except in the vicinity of
the symmetry-breaking edge, the curve of $\langle N_{2}\rangle_{\nu}$
versus $\mathcal{E}_{\nu}$ (red circles) overlaps with the curve
of $\langle N_{2}\rangle_{{\rm relax}}$ versus $E_{{\rm ini}}(\theta)$,
which implies dephasing is the underlying mechanism for the relaxation
process as described in Fig. \ref{fig:Projection}. The inset shows
the relationship between the initial state $\theta$ and the initial
energy $E_{{\rm ini}}$. We choose to present the result for $0\le\theta\le\pi/2$
out of the full range between $0$ and $\pi$. \label{fig:N2vsTheta}}
\end{figure}

\begin{figure}
\includegraphics[width=0.98\columnwidth]{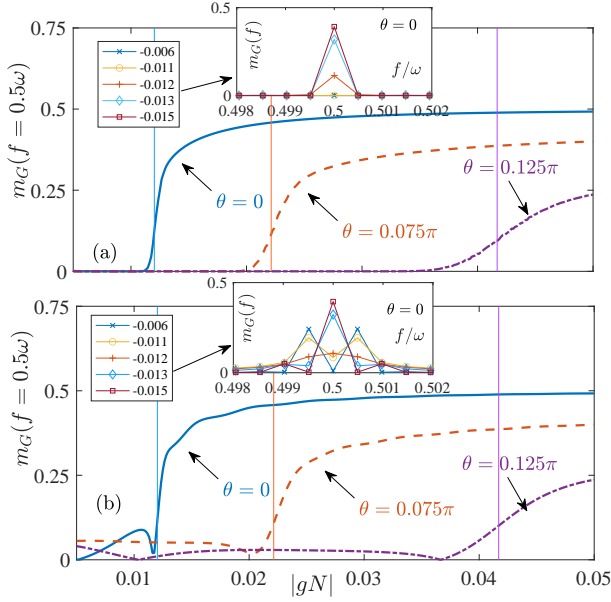}\caption{(a) Subharmonic response $|m_{G}(f=0.5\omega)|$ of the steady state
(at $t=k_{0}T$ to $(k_{0}+2048)T$ with $k_{0}=10^{5}$) as a function
of $gN$ for different initial \textcolor{black}{states $|\{\theta,\varphi=0\}\rangle$
defined in Eq. (\ref{eq:theta_state})} with $\theta=0,\ 0.075\pi$
and $0.125\pi$ shown by the blue solid, red dashed and purple dash-dotted
curves, respectively. The thin vertical lines indicate the critical
interaction $|g_{c}N|$. The inset shows the frequency response in
the window near $0.5\omega$ for different $gN$ indicated in the
legend. (b)\textcolor{black}{{} }The same as (a) for the transient state
at $t=0T$ to $2048T$. \label{fig:SubHarm}}
\end{figure}

A particularly useful observable to illustrate the importance of the
symmetry-breaking edge is $\langle N_{2}\rangle_{\nu}$ and $\langle N_{1}\rangle_{\nu}=N-\langle N_{2}\rangle_{\nu}$,
where $\langle O\rangle_{\nu}\equiv\left\langle \nu\right|\hat{O}\left|\nu\right\rangle $
and 
\begin{equation}
\langle N_{2}\rangle_{\nu}=\begin{cases}
\langle N_{2}\rangle_{\nu}^{(U)}\ {\rm or}\ \langle N_{2}\rangle_{\nu}^{(L)} & \mathcal{E}_{\nu}<\mathcal{E}_{{\rm edge}}\\
\langle N_{2}\rangle_{\nu}^{(S)} & \mathcal{E}_{\nu}\ge\mathcal{E}_{{\rm edge}}
\end{cases}.\label{eq:N2nu}
\end{equation}
For symmetry-unbroken states, the permutation symmetry between modes
ensures that $\langle N_{1}\rangle_{\nu}^{(S)}=\langle N_{2}\rangle_{\nu}^{(S)}=N/2$.
On the other hand, for symmetry-broken states, we have $\langle N_{1}\rangle_{\nu}^{(U)}=\langle N_{2}\rangle_{\nu}^{(L)}$
and $\langle N_{1}\rangle_{\nu}^{(L)}=\langle N_{2}\rangle_{\nu}^{(U)}$,
but in general $\langle N_{2}\rangle_{\nu}^{(U)}\ne\langle N_{2}\rangle_{\nu}^{(L)}\ne N/2$.
Without loss of generality, we denote $\langle N_{2}\rangle_{\nu}^{(U)}>\langle N_{2}\rangle_{\nu}^{(L)}$
(hence the U- and L-branch). Figures \ref{fig:Edge} (b), (d) and
(f) show $\langle N_{2}\rangle_{\nu}$ as a function of $\mathcal{E}_{\nu}$,
with initial state $|0,N\rangle$ and $N=1000$. The black dash-dotted
vertical line indicates the symmetry-broken edge, which illustrates
the onset of bifurcation of $\langle N_{2}\rangle_{\nu}$ as a function
of $\mathcal{E}_{\nu}$. We also show $\langle N_{2}\rangle_{{\rm relax}}$
by the purple pentagram symbol and the initial energy $E_{{\rm ini}}=\langle\Theta(0)|\hat{h}_{0}|\Theta(0)\rangle$
by the dashed vertical line. Numerically, $\langle N_{2}\rangle_{{\rm relax}}$
is calculated by averaging $\langle N_{2}\rangle_{{\rm relax}}=\sum_{k=k_{0}}^{k_{0}+K-1}\langle N_{2}(kT)\rangle/K$
over a time-window of $K=2048$ driving periods at $k_{0}T=10^{4}T$.
We can see that the relaxation value lies on the $\langle N_{2}\rangle_{\nu}-\mathcal{E}_{\nu}$
curve of the corresponding branches, which can be understood via dephasing.
In general, for an initial state $|\Theta(0)\rangle=\sum_{\nu}c_{\nu}|\nu\rangle$,
the time-evolution of an observable is given by $\langle O\rangle(t)=\sum_{\mu\nu}c_{\mu}^{*}c_{\nu}\langle\nu|O|\mu\rangle e^{-i(\mathcal{E}_{\nu}-\mathcal{E}_{\mu})t}$.
Typically, $c_{\nu}$ only has noticeable values around a small energy
window close to the initial energy. If there is no degeneracy, after
long enough time, dephasing leads to $\langle O\rangle_{{\rm relax}}=\sum_{\nu}p_{\nu}\langle O\rangle_{\nu}={\rm Tr}[\rho_{d}\hat{O}]$,
where $\rho_{d}=\sum_{\nu}p_{\nu}|\nu\rangle\langle\nu|$ is sometimes
called the diagonal ensemble, and $p_{\nu}=|c_{\nu}|^{2}$ is the
projection probability of the initial state to the $\nu$'s eigenstate.
Essentially, the contributions for $\mu\neq\nu$ cancel out for large
$t$, as the phase factors in $\langle O\rangle(t)$ become more random.
The initial energy can also be given by the diagonal ensemble $E_{{\rm ini}}=\sum_{\nu}p_{\nu}\mathcal{E_{\nu}}={\rm Tr}[\rho_{d}\hat{h}_{0}]$.
When the initial energy is well above the broken edge, $\langle N_{2}\rangle_{\nu}=N/2$
gives $\langle N_{2}\rangle_{{\rm relax}}=N/2$ as shown in Fig. \ref{fig:Projection}
(a). When the initial energy is below the broken edge, as shown in
Fig. \ref{fig:Projection} (b) and (c), we find that for initial state
$|0,N\rangle$, only the U-branch has noticeable projections, since
the degeneracy is lifted in a single branch, the dephasing formula
still works, and gives $\langle N_{2}\rangle_{{\rm relax}}=\sum_{\nu}p_{\nu}\langle N_{2}\rangle_{\nu}$.

To further explore the corresponding relation between the initial
energy and the relaxation value, we investigate the case with an initial
state $|\Theta(0)\rangle=|\{\theta,\varphi\}\rangle$ defined in Eq.
(\ref{eq:theta_state}) with all bosons in a mode given by $\Lambda_{2}=\sin\theta e^{i\varphi}\Phi_{1}+\cos\theta\Phi_{2},$
which has initial energy $E_{{\rm ini}}(\theta,\varphi)=\langle\{\theta,\varphi\}|\hat{h}_{0}|\{\theta,\varphi\}\rangle$.
Since this initial state represents all atoms occupying the same single-particle
state, the initial energy reduces to a mean-field expression 
\begin{equation}
\begin{aligned}\frac{E_{{\rm ini}}(\theta,\varphi)}{N}\approx & J\sin2\theta\cos(\varphi)+gN\left[\frac{u_{I}}{2}+\right.\\
 & \left.u_{T}\sin2\theta\cos(\varphi)+\frac{(2u_{N}-u_{I})}{4}\sin^{2}2\theta\right].
\end{aligned}
\label{eq:Eini}
\end{equation}
Figure \ref{fig:N2vsTheta} shows $\langle N_{2}\rangle_{{\rm relax}}$
as a function of the initial energy $E_{{\rm ini}}(\theta,\varphi=0)$
for different initial states $|\{\theta,\varphi=0\}\rangle$, and
$\langle N_{2}\rangle_{\nu}$ as a function of $\mathcal{E}_{\nu}$
for $gN=-0.1$. As one can see, in the deep regime in all branches
(L, U and S), these two curves overlap, implying the dephasing mechanism
leads to relaxation except very close to the critical point where
finite-size effects become important. The inset shows the corresponding
relationship between the initial state $\theta$ and the initial energy
$E_{{\rm ini}}(\theta,\varphi=0)$. We also note that the initial
states considered here are initial states with no multi-mode entanglement,
as these are the initial states of interest for spontaneous symmetry-breaking
physics.

\begin{figure}
\includegraphics[width=0.98\columnwidth]{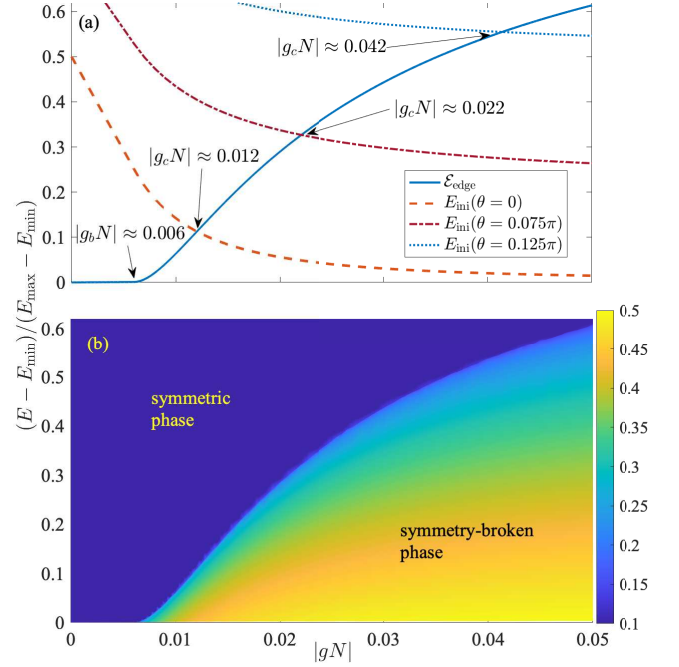}\caption{(a) $\mathcal{E}_{{\rm edge}}$ (blue solid curve) as a function of
$|gN|$ and the initial energy $E_{{\rm ini}}(\theta)\equiv E_{{\rm ini}}(\theta,\varphi=0)$
as a function of $|gN|$ for different initial states $|\Theta(0)\rangle=|\{\theta,\varphi=0\}\rangle$
with $\theta=0,\ 0.075\pi$ and $0.125\pi$, (red dashed, purple dash-dotted
and light blue dotted, respectively). The initial energy curve crosses
the edge at the critical $|g_{c}N|$ shown in Fig. \ref{fig:SubHarm}.
(b) Subharmonic response $|m_{G}(\omega/2)|$ (colour bar) as a function
of normalized initial energy $E_{{\rm ini}}$ and $gN$ for all initial
states $|\{\theta,\varphi=0\}\rangle$ {[}with $\theta\in[0,\pi)${]},
indicating the symmetry-broken and symmetric phase. This phase diagram
also applies to other possible initial product states $|\{\theta,\varphi\}\rangle$,
with the exception of some regions where the condition $\varphi\protect\ne0$
limits the range of $E_{{\rm ini}}(\theta,\varphi)$ that can be accessed
(see Appendix \ref{sec:mf} for details regarding $\varphi\protect\ne0$
cases). \label{fig:phasediagram}}
\end{figure}

In Fig. \ref{fig:SubHarm} (a), we investigate the subharmonic response
$|m_{G}(\omega/2)|$ of the steady state as a function of $|gN|$
for different initial states $|\Theta(0)\rangle=|\{\theta,\varphi=0\}\rangle$,
which changes from $0$ to a finite value abruptly around a critical
$|g_{c}N|$ indicated by the thin vertical lines. These critical values
for different $\theta$ can be obtained by equating $E_{{\rm ini}}(\theta,\varphi=0)$
and $\mathcal{E}_{{\rm edge}}$ as shown in Fig. \ref{fig:phasediagram}
(a). The inset of Fig. \ref{fig:SubHarm} (a) shows the frequency
response $|m_{G}(f)|$ for $f$ close to $\omega/2$ and the initial
state $|\Theta(0)\rangle=|\{\theta=0,\varphi=0\}\rangle=|0,N\rangle$.
We can see that a single sharp peak appears abruptly when $|gN|\geq|g_{c}N|$.
However, in realistic experiments, one might not be able to access
such very long evolution times. In Fig. \ref{fig:SubHarm} (b), we
show that the transient behavior at relatively short times can be
regarded as a precurser. One can see that, while not as smooth and
clean as Fig. \ref{fig:SubHarm} (a), $|m_{G}(\omega/2)|$ still shows
an abrupt change to a much larger value near $|g_{c}N|$. In the inset,
one can see that the frequency response of the transient states for
$\theta=0$ at short times shows a double-peak structure for $|gN|<|g_{c}N|$
and which changes to a single-peak structure for $|gN|\geq|g_{c}N|$,
the same as the observation in \citep{WangNJP2021}. In the steady-state
regime, we now have the intriguing situation where, as the magnitude
of the interaction is increased from that just below the critical
interaction $|g_{c}N|$ to just above $|g_{c}N|$, the period of the
bouncing atom cloud changes from period $T$ to period $2T$, to form
a DTC. Although the numerical results presented here are calculated
for a finite $N=1000$, we believe the conclusions remain valid in
the thermodynamic limit ($N\rightarrow\infty$ and finite $gN$).
An analysis of the finite size effect is given in Appendix \ref{sec:FiniteSize}.

The whole phase diagram for different initial energy and interaction
strength is summarized in Fig. \ref{fig:phasediagram}. The solid
line in Fig. \ref{fig:phasediagram} (a) shows the normalized $\mathcal{E}_{{\rm edge}}$
, which agrees with the phase boundary indicated by the subharmonic
response $m_{G}(f=\omega/2)$ in Fig. \ref{fig:phasediagram} (b).
The dashed, dash-dotted and dotted curves in Fig. \ref{fig:phasediagram}
(a) show the initial energy $E_{{\rm ini}}(\theta,\varphi=0)$ as
a function of $|gN|$ for different initial states $|\Theta(0)\rangle=|\{\theta,\varphi=0\}\rangle$
defined in Eq. (\ref{eq:theta_state}) with $\theta=0,\ 0.075\pi$
and $0.125\pi$, respectively. The initial energy curve crosses the
edge at the critical $|g_{c}N|$ shown in Fig. \ref{fig:SubHarm}.
For a general initial product state $|\{\theta,\varphi\}\rangle$,
the critical $|g_{c}N|$ as a function of $\theta$ for a given $\varphi$
is shown in Fig. \ref{fig:gcN}. (See Appendix \ref{sec:mf} for details.)
This shows that the onset of DTC formation depends on the choice of
initial state $(\theta,\varphi)$ through the initial energy $E_{{\rm ini}}(\theta,\varphi)$,
as well as on the system parameters $J$, $N$. Here, the critical
$|g_{c}N|$ for the onset of DTC formation is determined by equating
the initial energy $E_{{\rm ini}}(\theta,\varphi)$ with the $E_{{\rm edge}}$.

These phase diagrams indicate that the symmetry of the steady state
characterized by the subharmonic response $|m_{G}(\omega/2)|$ is
determined by whether the initial energy is above or below the symmetry-breaking
edge of the eigenenergy spectrum for a given interaction strength
$gN$. The symmetry-breaking edge thus gives the abrupt phase transition
boundary. This phase transition boundary is valid for any general
initial product states $|\{\theta,\varphi\}\rangle$ (see Appendix
\ref{sec:mf} for details).

\begin{figure}
\includegraphics[width=0.98\columnwidth]{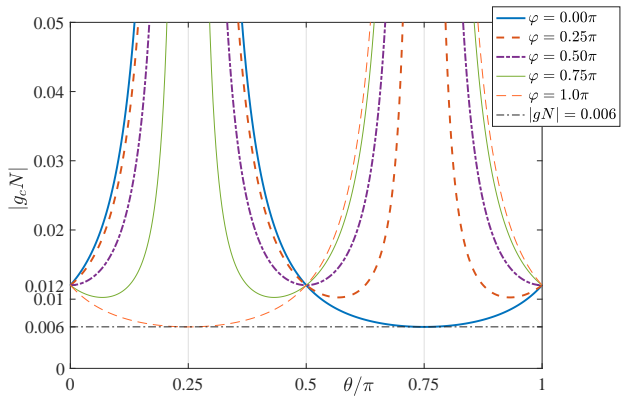}\caption{$|g_{c}N|$ for different initial product states $|\{\theta,\varphi\}\rangle$.
The minimum possible $|g_{c}N|\approx0.006$ corresponds to $\{\theta,\varphi\}=\{0.75\pi,0\}$
(or $\{0.25\pi,\pi\}$ which corresponds to the same physical state).
The condition $\theta=0$ ($\theta=\pi/2$) corresponds to the initial
state being $|0,N_{2}=N\rangle$ ($|N_{1}=N,0\rangle$), where $\varphi$
is irrelevant and $|g_{c}N|\approx0.012$. \label{fig:gcN}}
\end{figure}

This $\mathbb{Z}_{2}$ symmetry breaking is associated with the time-translational
symmetry breaking in the time crystal. Denoting the elements of the
eigenstate vector $\vec{f}_{\nu}$ as $f_{n}^{(\nu)}$, the eigenstates
can be written as $\left|\nu;t\right\rangle =\sum_{n}f_{n}^{(\nu)}\left|n,N-n;t\right\rangle $.
The permutation operator $\hat{P}_{12}\left|n,N-n;t\right\rangle =\left|N-n,n;t\right\rangle $
transforms the eigenstates as

\begin{equation}
\begin{alignedat}{1}\hat{P}_{12}\left|\nu;t\right\rangle  & =\sum_{n}f_{n}^{(\nu)}\left|N-n,n;t\right\rangle \\
 & =\sum_{n}f_{n}^{(\nu)}\left|n,N-n;t+T\right\rangle =\left|\nu;t+T\right\rangle .
\end{alignedat}
\end{equation}
For $\mathbb{Z}_{2}$ symmetry-unbroken eigenstates, we also have
$\hat{P}_{12}\left|\nu;t\right\rangle _{S}=\pm\left|\nu;t\right\rangle _{S}$,
which leads to $\left|\nu;t+T\right\rangle _{S}=\pm\left|\nu;t\right\rangle _{S}$.
Therefore, these eigenstates are $T$-periodic, and satisfy the same
discrete time-translational symmetry as the Hamiltonian. In contrast,
for a pair of $\mathbb{Z}_{2}$ symmetry-broken approximate eigenstates,
we have $\hat{P}_{12}\left|\nu;t\right\rangle _{U}=\left|\nu;t\right\rangle _{L}=\left|\nu;t+T\right\rangle _{U}\ne\left|\nu;t\right\rangle _{U}$.
These eigenstates therefore have a period of $2T$ instead of $T$,
which breaks the discrete time-translational symmetry of the Hamiltonian.
The eigenstates of $\hat{h}_{0}$ can be regarded as an approximation
of the many-body Floquet states in Eq. (9) with period $2T$. Therefore,
the near-degeneracy of $|\nu\rangle_{L}$ and $|\nu\rangle_{U}$ in
the symmetry-breaking regime plays the same role as the nearly $\pi$-pairing
of $T$-Floquet states in MBL and many-body quantum scar DTCs \citep{SachaBook2020}.

\section{Summary}

We have derived a two-mode model from standard quantum field theory
to study discrete time-translational symmetry breaking and the formation
of DTCs in a Bose-Einstein condensate bouncing resonantly on an oscillating
mirror. The validity of this simple many-body model has been investigated
by comparing with previous multimode phase-space many-body calculations
based on the TWA approach \citep{WangNJP2021}. The greatly reduced
computational times using this simple many-body model allows us to
study the long-time dynamical evolution of the many-body system. Using
two Wannier modes constructed from two single-particle Floquet states,
the dynamical evolution has been studied both via a fully time-dependent
Floquet Hamiltonian (MBF) and by using a time-independent Hamiltonian
based on a high frequency expansion (HFE). In the HFE approach, the
Hamiltonian is equivalent to the Lipkin-Meshkov-Glick model \citep{Ribeiro2008PRE,Fabrizio2012PRB,KurosNJP2020}.
The main initial state chosen has all bosons in the Wannier mode that
closely resembles the condensate mode for a BEC in a harmonic trap
treated in our TWA approach. However, a wide variety of initial states
based on the two Wannier modes has also been studied. A new criterion
for demonstrating the periodicity at stroboscopic times has been developed
which involves the mean number of bosons in each Wannier mode.

We find that the evolution investigated in previous studies in the
time-window out to about 2000 driving periods actually involves ``short-time''
transient phenomena though DTC formation is still
shown if the inter-boson interaction is strong enough. The two-mode
compares well with the TWA approach in regard to the critical $|g_{c}N|$
for DTC formation. However the TWA treatment is needed to verify that
quantum depletion to other modes is negligible. After much longer
evolution times, initial states with no long-range correlations relax
to a steady state and eventually show short-lived quantum revivals.
In the steady state the mean boson number in each Wannier mode demonstrates
that stroboscopic DTC behavior occurs for the same interaction regime
found in the previous TWA calculations, the critical value for DTC
formation with $2T$ periodicity being about $g_{c}N=-0.012$ for
the parameters considered and the initial state $|\{\theta=0,\varphi=0\}\rangle$,
which is the easiest to prepare in an experiment. However, the two-mode
theory now more clearly shows that in the steady-state regime, for
the initial state $|\{\theta=0,\varphi=0\}\rangle$ and for smaller
$|gN|$, only $T$ periodicity occurs.

For a general initial product state condition $\{\theta,\varphi\}$,
the magnitude of the critical value $|g_{c}N|$ can be as low as $-0.006$
(see Fig. \ref{fig:gcN} and Appendix \ref{sec:mf}). The long-time
behavior can be understood via the many-body Floquet quasi-eigenenergy
spectrum of the two-mode model. For sufficiently strong interaction,
a symmetry-breaking edge appears in the spectrum, where all quasi-eigenstates
below the edge are symmetry-breaking while those above the edge are
symmetric. The position of the edge is found to depend on the boson
number and the inter-mode tunnelling rate, and gives $g_{c}N$ as
a function of $E_{{\rm ini}}$ without explicit dependence on initial
details $\{\theta,\varphi\}$. Finally, a phase diagram showing regions
of symmetry-broken and symmetric phases for differing initial energies
and interaction strengths summarises the subharmonic response results.
Here, we now allow for initial states where all bosons occupy a mode
which is a linear combination of the two Wannier modes parameterized
by $\{\theta,\varphi\}$. Our results predict that in the steady-state
regime, after about $50,000$ driving periods (for the parameters
considered), as the magnitude of the interaction is increased from
just below to just above the critical interaction strength the period
of the bouncing atom cloud changes abruptly from the driving period
$T$ to period $2T$, to form a discrete time crystal. The
present two-mode theory approach predicts that the discrete time crystal
survives for times out to 250,000 driving periods.  However, 
after astronomically long time, the escape of atoms to other modes beyond our
two-mode model might eventually occur, leading to a finite lifetime of our
long-lived DTC.

\section{ACKNOWLEDGEMENTS}

J.W acknowledges support from an ARC DECRA Grant no. DE180100592).
The project is supported by an ARC Discovery Project grant (Grant
no. DP190100815). K.S. acknowledges support of the National Science
Centre Poland via Project 2018/31/B/ST2/00349.

\appendix

\section{Derivations}

In this Appendix we derive some of the key equations used in the main
body of the paper.

\subsection{Derivation of Equations (\ref{eq:EvolutionEq}), (\ref{eq:H2tmode})\label{subsec:Derivation1}}

Substituting the two-mode expansion Eq. (\ref{eq:TwoModeExpansion})
of the field operators into the Hamiltonian in Eq. (\ref{eq:Hamiltonian})
gives

\begin{equation}
\hat{H}_{2m}=\sum_{i,j=1,2}E_{i,j}\hat{a}_{i}^{\dagger}\hat{a}_{j}+\frac{g}{2}\sum_{i,j,k,l=1,2}U_{i,j,k,l}\hat{a}_{i}^{\dagger}\hat{a}_{j}^{\dagger}\hat{a}_{k}\hat{a}_{l},
\end{equation}
where 
\begin{equation}
E_{i,j}=\int dz\Phi_{i}(z,t)^{*}H_{sp}\Phi_{j}(z,t)
\end{equation}
\begin{equation}
U_{i,j,k,l}=\int dz\Phi_{i}(z,t)^{*}\Phi_{j}(z,t)^{*}\Phi_{k}(z,t)\Phi_{l}(z,t).
\end{equation}

Using Eq. (\ref{eq:WannierStates}) for the Wannier modes and Eq.
(\ref{eq:Floquet_sp}) for the Floquet modes we then find that 
\begin{equation}
\begin{array}{l}
E_{1,1}=\left(\epsilon_{1}+\epsilon_{2}-\hbar\omega/2\right)/2-\hbar D_{1,1}=\epsilon_{1,1}-\hbar D_{1,1}\\
E_{1,2}=\left(\epsilon_{1}-\epsilon_{2}+\hbar\omega/2\right)/2-\hbar D_{1,2}=\epsilon_{1,2}-\hbar D_{1,2}\\
E_{2,1}=\left(\epsilon_{1}-\epsilon_{2}+\hbar\omega/2\right)/2-\hbar D_{2,1}=\epsilon_{2,1}-\hbar D_{2,1}\\
E_{2,2}=\left(\epsilon_{1}+\epsilon_{2}-\hbar\omega/2\right)/2-\hbar D_{2,2}=\epsilon_{2,2}-\hbar D_{2,2},
\end{array}
\end{equation}
where 
\begin{equation}
D_{j,i}=-i\int dz\Phi_{j}(z,t)^{*}\frac{\partial}{\partial t}\Phi_{i}(z,t)=D_{i,j}^{*}
\end{equation}

\begin{equation}
\epsilon_{1,1}=\epsilon_{2,2}=\left(\epsilon_{1}+\epsilon_{2}-\hbar\omega/2\right)/2=\epsilon
\end{equation}

\begin{equation}
\epsilon_{1,2}=\epsilon_{2,1}=\left(\epsilon_{1}-\epsilon_{2}+\hbar\omega/2\right)/2=J.
\end{equation}
Hence the Hamiltonian is given by 
\begin{equation}
\hat{H}_{2m}=\sum_{i,j=1,2}\left(\epsilon_{i,j}-\hbar D_{i,j}\right)\hat{a}_{i}^{\dagger}\hat{a}_{j}+\frac{g}{2}\sum_{i,j,k,l=1,2}U_{i,j,k,l}\hat{a}_{i}^{\dagger}\hat{a}_{j}^{\dagger}\hat{a}_{k}\hat{a}_{l}.
\end{equation}
We now expand the quantum state as

\begin{equation}
|\Theta(t)\rangle=\sum_{n}B_{n}(t)|n,N-n\rangle,
\end{equation}
where $\left\vert n_{1},n_{2}\right\rangle $ are Fock states given
by Eq. (6), with $n_{1}$, $n_{2}$ bosonic atoms in modes $\Phi_{1}$,
$\Phi_{2}$, respectively. Note that these basis states are time-dependent,
as are the amplitudes $B_{n}(t)$.

Substituting into the time-dependent Schr{ö}dinger equation then
gives 
\begin{equation}
\begin{aligned}\sum_{m}B_{m}(t)\hat{H}_{2m}|m,N-m\rangle & =i\hbar\sum_{n}\left(\frac{\partial}{\partial t}B_{n}(t)\right)|n,N-n\rangle\\
 & +i\hbar\sum_{m}B_{m}(t)\left(\frac{\partial}{\partial t}|m,N-m\rangle\right),
\end{aligned}
\end{equation}
so taking the scalar product with $\left\langle n,N-n\right\vert $
on each side gives

\begin{equation}
\begin{aligned}i\hbar\left(\frac{\partial}{\partial t}B_{n}(t)\right) & =\sum_{m}B_{m}(t)\left\langle n,N-n\left|\hat{H}_{2m}\right|m,N-m\right\rangle \\
 & -i\hbar\sum_{m}B_{m}(t)\langle n,N-n|\left(\frac{\partial}{\partial t}|m,N-m\rangle\right).
\end{aligned}
\label{eq:AmpB}
\end{equation}
We can eliminate the $D_{i,j}$ terms by using the expansion for the
time-independent field creation operator to first derive an equation
for the time derivative of the Wannier mode creation operators

\begin{equation}
\begin{aligned}0 & =\sum_{i}\Phi_{i}(z,t)^{*}\frac{\partial}{\partial t}\hat{a}_{i}^{\dagger}+\sum_{j}\frac{\partial}{\partial t}\Phi_{j}(z,t)^{*}\hat{a}_{j}^{\dagger}\\
\frac{\partial}{\partial t}\hat{a}_{i}^{\dagger} & =-\sum_{j}\int dz\left(\Phi_{i}(z,t)\frac{\partial}{\partial t}\Phi_{j}(z,t)^{*}\right)\hat{a}_{j}^{\dagger}=i\sum_{j}D_{j,i}\hat{a}_{j}^{\dagger}.
\end{aligned}
\end{equation}
Hence the time derivative of the basis states is 
\begin{widetext}
\begin{equation}
\begin{aligned}\frac{\partial}{\partial t}\left|n_{1},n_{2}\right\rangle = & \left\{ \left(\frac{\partial}{\partial t}\left(\hat{a}_{1}^{\dagger}\right)^{n_{1}}\right)\left(\hat{a}_{2}^{\dagger}\right)^{n_{2}}+\left(\hat{a}_{1}^{\dagger}\right)^{n_{1}}\left(\frac{\partial}{\partial t}\left(\hat{a}_{2}^{\dagger}\right)^{n_{2}}\right)\right\} |0,0\rangle/\left(\sqrt{n_{1}!}\sqrt{n_{2}!}\right)\\
= & \left\{ n_{1}\left(\hat{a}_{1}^{\dagger}\right)^{n_{1}-1}\left(\frac{\partial}{\partial t}\left(\hat{a}_{1}^{\dagger}\right)\right)\left(\hat{a}_{2}^{\dagger}\right)^{n_{2}}+\left(\hat{a}_{1}^{\dagger}\right)^{n_{1}}n_{2}\left(\hat{a}_{2}^{\dagger}\right)^{n_{2}-1}\left(\frac{\partial}{\partial t}\left(\hat{a}_{2}^{\dagger}\right)\right)\right\} |0,0\rangle/\left(\sqrt{n_{1}!}\sqrt{n_{2}!}\right)\\
= & \left\{ n_{1}\left(\hat{a}_{1}^{\dagger}\right)^{n_{1}-1}i\sum_{j}D_{j,1}\hat{a}_{j}^{\dagger}\left(\hat{a}_{2}^{\dagger}\right)^{n_{2}}+\left(\hat{a}_{1}^{\dagger}\right)^{n_{1}}n_{2}\left(\hat{a}_{2}^{\dagger}\right)^{n_{2}-1}i\sum_{j}D_{j,2}\hat{a}_{j}^{\dagger}\right\} |0,0\rangle/\left(\sqrt{n_{1}!}\sqrt{n_{2}!}\right)\\
= & i\left\{ n_{1}\left(\hat{a}_{1}^{\dagger}\right)^{n_{1}}D_{1,1}\left(\hat{a}_{2}^{\dagger}\right)^{n_{2}}+\left(\hat{a}_{1}^{\dagger}\right)^{n_{1}+1}n_{2}\left(\hat{a}_{2}^{\dagger}\right)^{n_{2}-1}D_{1,2}\right\} |0,0\rangle/\left(\sqrt{n_{1}!}\sqrt{n_{2}!}\right)\\
 & +i\left\{ n_{1}D_{2,1}\left(\hat{a}_{1}^{\dagger}\right)^{n_{1}-1}\left(\hat{a}_{2}^{\dagger}\right)^{n_{2}+1}+n_{2}D_{2,2}\left(\hat{a}_{1}^{\dagger}\right)^{n_{1}}\left(\hat{a}_{2}^{\dagger}\right)^{n_{2}}\right\} |0,0\rangle/\left(\sqrt{n_{1}!}\sqrt{n_{2}!}\right)\\
= & i\left\{ n_{1}D_{1,1}\left|n_{1},n_{2}\right\rangle +\sqrt{n_{1}+1}\sqrt{n_{2}}D_{1,2}\left|n_{1}+1,n_{2}-1\right\rangle \right\} \\
 & +i\left\{ \sqrt{n_{1}}\sqrt{n_{2}+1}D_{2,1}\left|n_{1}-1,n_{2}+1\right\rangle +n_{2}D_{2,2}\left|n_{1},n_{2}\right\rangle \right\} \\
= & i\left\{ D_{1,1}\hat{a}_{1}^{\dagger}\hat{a}_{1}+D_{1,2}\hat{a}_{1}^{\dagger}\hat{a}_{2}+D_{2,1}\hat{a}_{2}^{\dagger}\hat{a}_{1}+D_{2,2}\hat{a}_{2}^{\dagger}\hat{a}_{2}\right\} \left|n_{1},n_{2}\right\rangle \\
= & \left(\sum_{i,j}D_{i,j}\hat{a}_{i}^{\dagger}\hat{a}_{j}\right)\left|n_{1},n_{2}\right\rangle ,
\end{aligned}
\end{equation}
\end{widetext}

where we have used the result that the $\partial_{t}\hat{a}_{i}^{\dagger}$
commute with any $\hat{a}_{j}^{\dagger}$.

Hence on substituting for $\left(\frac{\partial}{\partial t}\left\vert m,N-m\right\rangle \right)$
in Eq. (\ref{eq:AmpB}) we see that the $D_{i,j}$ terms cancel out
leaving 
\begin{equation}
\begin{array}{r}
i\hbar\left(\frac{\partial}{\partial t}B_{n}(t)\right)=\sum_{m}B_{m}(t)\langle n,N-n|\left(\sum_{i,j=1,2}\epsilon_{i,j}\hat{a}_{i}^{\dagger}\hat{a}_{j}\right.\\
\left.+\frac{g}{2}\sum_{i,j,k,l=1.2}U_{i,j,k,l}\hat{a}_{i}^{\dagger}\hat{a}_{j}^{\dagger}\hat{a}_{k}\hat{a}_{l}\right)|m,N-m\rangle
\end{array}.
\end{equation}
If we write $B_{n}(t)=b_{n}(t)\exp(-i\epsilon Nt/\hbar)$ we then
find that the terms involving $\epsilon$ cancel out leaving 
\begin{equation}
\begin{aligned}i\hbar\left(\frac{\partial}{\partial t}b_{n}(t)\right)= & \sum_{m}b_{m}(t)\langle n,N-n|\left[J\left(\hat{a}_{1}^{\dagger}\hat{a}_{2}+\hat{a}_{2}^{\dagger}\hat{a}_{1}\right)\right.\\
+ & \frac{g}{2}\sum_{i,j,k,l=1,2}U_{i,j,k,l}\left.\hat{a}_{i}^{\dagger}\hat{a}_{j}^{\dagger}\hat{a}_{k}\hat{a}_{l}\right]|m,N-m\rangle
\end{aligned}
,
\end{equation}
so we now have 
\begin{equation}
\sum_{m}\langle n,N-n|(\hat{\mathcal{H}})|m,N-m\rangle b_{m}(t)-i\hbar\left(\frac{\partial}{\partial t}b_{n}(t)\right)=0\label{eq:FinalAmpl}
\end{equation}
\begin{equation}
\left(\underline{\mathcal{H}}-i\hbar\frac{\partial}{\partial t}\right)\vec{b}=0,\label{eq:FinalAmpMatrixForm}
\end{equation}
where 
\begin{equation}
\hat{\mathcal{H}}=J\left(\hat{a}_{1}^{\dagger}\hat{a}_{2}+\hat{a}_{2}^{\dagger}\hat{a}_{1}\right)+\frac{g}{2}\sum_{i,j,k,l=1,2}U_{i,j,k,l}\hat{a}_{i}^{\dagger}\hat{a}_{j}^{\dagger}\hat{a}_{k}\hat{a}_{l}\label{eq:EffectiveHamilt}
\end{equation}
and the quantum state is now given by 
\begin{equation}
|\Theta(t)\rangle=\exp(-i\epsilon Nt/\hbar)\sum_{n}b_{n}(t)|n,N-n\rangle.
\end{equation}
The phase factor $\exp(-i\epsilon Nt/\hbar)$ is not physically important.
Thus, Eqs. (\ref{eq:FinalAmpMatrixForm}), (\ref{eq:EffectiveHamilt})
are equivalent to Eqs. (7), (8) in the main part of the paper after
introducing the matrix $\underline{\mathcal{H}}$ for $\hat{\mathcal{H}}$
and a column vector $\vec{b}$ for the amplitudes $b_{n}(t)$ via
$\underline{\mathcal{H}}_{nm}=\left\langle n,N-n\right\vert \hat{\mathcal{H}}\left\vert m,N-m\right\rangle $
and $\vec{b}_{n}=b_{n}(t)$.

\subsection{Derivation of Equation (\ref{eq:h0})\label{subsec:Derivation2}}

At this stage the HFE approximation has been made, so that $\hat{\mathcal{H}}$
is replaced by $\hat{h}_{0}$ given by 
\begin{equation}
\hat{h}_{0}=J\left(\hat{a}_{1}^{\dagger}\hat{a}_{2}+\hat{a}_{2}^{\dagger}\hat{a}_{1}\right)+\frac{g}{2}\sum_{i,j,k,l=1,2}\bar{U}_{i,j,k,l}\hat{a}_{i}^{\dagger}\hat{a}_{j}^{\dagger}\hat{a}_{k}\hat{a}_{l}
\end{equation}
\begin{equation}
\bar{U}_{i,j,k,l}=\frac{1}{2T}\int dtU_{i,j,k,l},
\end{equation}
which now means that all the coefficients in $\hat{h}_{0}$ are time-independent.

Many of the $16$ coefficients $\bar{U}_{i,j,k,l}$ are inter-related.
Firstly, from the definition of $U_{i,j,k,l}$ we see that

\begin{equation}
\bar{U}_{i,j,k,l}=\bar{U}_{j,i,k,l}=\bar{U}_{i,j,l,k},
\end{equation}
which leads to

\begin{equation}
\begin{array}{l}
\bar{U}_{11,12}=\bar{U}_{11,21}\quad\bar{U}_{22,12}=\bar{U}_{22,21}\\
\bar{U}_{12,11}=\bar{U}_{21,11}\quad\bar{U}_{12,22}=\bar{U}_{21,22}\\
\bar{U}_{12,12}=\bar{U}_{12,21}=\bar{U}_{21,12}=\bar{U}_{21,21}.
\end{array}
\end{equation}
In addition to these $12$ coefficients, there are $4$ more, namely
$\bar{U}_{11,22}$, $\bar{U}_{22,11}$, $\bar{U}_{11,11}$ and $\bar{U}_{22,22}$.

We can show more inter-relationships by expressing the Wannier states
by introducing the notation $\eta(t)=\exp(-i\pi t/T)$ and $s_{i}=1$
for Wannier mode $\Phi_{1}$ and $s_{i}=-1$ for Wannier mode $\Phi_{2}$.
Thus $\Phi_{i}(z,t)=a(\phi_{1}(z,t)+s_{i}\eta(t)\phi_{2}(z,t))$,
where $a=1/\sqrt{2}$. We can then divide the time interval $0$ to
$2T$ in the definition for $\bar{U}_{i,j,k,l}$ into time intervals
$0$ to $T$ and $T$ to $2T$, and then make use of the properties
$\phi_{i}(z,t+T)=\phi_{i}(z,t)$ and $\eta(t+T)=-\eta(t)$ to convert
the integral from $T$ to $2T$ back into an integral from $0$ to
$T$. This gives the following expression for $\bar{U}_{i,j,k,l}$
\begin{equation}
\begin{aligned}\bar{U}_{i,j,k,l}= & \frac{a^{4}}{2T}\int_{0}^{T}dt\int dz\left(\begin{array}{c}
\left(\phi_{1}^{*}+s_{i}\eta^{*}\phi_{2}^{*}\right)\left(\phi_{1}^{*}+s_{j}\eta^{*}\phi_{2}^{*}\right)\\
\times\left(\phi_{1}+s_{k}\eta\phi_{2}\right)\left(\phi_{1}+s_{l}\eta\phi_{2}\right)
\end{array}\right)\\
+ & \frac{a^{4}}{2T}\int_{0}^{T}dt\int dz\left(\begin{array}{c}
\left(\phi_{1}^{*}-s_{i}\eta^{*}\phi_{2}^{*}\right)\left(\phi_{1}^{*}-s_{j}\eta^{*}\phi_{2}^{*}\right)\\
\times\left(\phi_{1}-s_{k}\eta\phi_{2}\right)\left(\phi_{1}-s_{l}\eta\phi_{2}\right)
\end{array}\right),
\end{aligned}
\label{eq:NewFormUBar}
\end{equation}
where the $z$, $t$ dependence of the functions is left understood.
This enables us to establish more inter-relationships, as the second
term for one $\bar{U}_{i,j,k,l}$ is often the first term for another
$\bar{U}_{i,j,k,l}$, and vice-versa. Hence we find that 
\begin{equation}
\begin{array}{llc}
\bar{U}_{11,11} & =\bar{U}_{22,22} & \bar{U}_{11,22}=\bar{U}_{22,11}\\
\bar{U}_{22,21} & =\bar{U}_{11,12} & \bar{U}_{12,11}=\bar{U}_{21,22}.
\end{array}
\end{equation}

Also, we see from Eq. (\ref{eq:NewFormUBar}) that there is a further
relationship 
\begin{equation}
\bar{U}_{i,j,k,l}^{*}=\bar{U}_{k,l,i,j},
\end{equation}
which leads to

\begin{equation}
\bar{U}_{12,11}=\bar{U}_{11,12}^{*}=\bar{U}_{11,12}
\end{equation}
since we can show that 
\begin{equation}
\bar{U}_{11,12}=\frac{a^{4}}{T}\int_{0}^{T}dt\int dz\left|\phi_{1}+\eta\phi_{2}\right|^{2}\left(\left|\phi_{1}\right|^{2}-\left|\phi_{2}\right|^{2}\right)
\end{equation}
is real.

Also, we have 
\begin{equation}
\begin{array}{l}
\bar{U}_{12,12}=\frac{a^{4}}{T}\int_{0}^{T}dt\int dz\left(\left|\phi_{1}+\eta\phi_{2}\right|^{2}\mid\left(\phi_{1}-\left.\eta\phi_{2}\right|^{2}\right)\right.\\
\bar{U}_{11,22}=\frac{a^{4}}{T}\int_{0}^{T}dt\int dz\left(\begin{array}{c}
\left(\left|\phi_{1}\right|^{2}-\left|\phi_{2}\right|^{2}\right)^{2}\\
+\left(\eta\phi_{1}^{*}\phi_{2}-\eta^{*}\phi_{2}^{*}\phi_{1}\right)^{2}
\end{array}\right)\\
\bar{U}_{11,11}=\frac{a^{4}}{2T}\int_{0}^{T}dt\int dz\left(\left|\phi_{1}+\eta\phi_{2}\right|^{4}+\left|\phi_{1}-\eta\phi_{2}\right|^{4}\right),
\end{array}
\end{equation}
which are all real.

Hence, overall we have only $4$ independent coefficients which can
be listed as 
\begin{equation}
\begin{aligned}u_{T} & =\bar{U}_{11,12}=\bar{U}_{11,21}=\bar{U}_{22,12}=\bar{U}_{22,21}\\
 & =\bar{U}_{12,11}=\bar{U}_{21,11}=\bar{U}_{12,22}=\bar{U}_{21,22}\\
u_{N} & =\bar{U}_{12,12}=\bar{U}_{12,21}=\bar{U}_{21,12}=\bar{U}_{21,21}\\
u_{P} & =\bar{U}_{11,22}=\bar{U}_{22,11}\\
u_{I} & =\bar{U}_{11,11}=\bar{U}_{22,22}.
\end{aligned}
\end{equation}

This obviously enables the expression for $\hat{h}_{0}$ to be simplified.
Hence we have 
\begin{equation}
\begin{aligned}\hat{h}_{0}= & J\left(\hat{a}_{1}^{\dagger}\hat{a}_{2}+\hat{a}_{2}^{\dagger}\hat{a}_{1}\right)\\
 & +\frac{g}{2}u_{T}\left(\begin{array}{c}
\hat{a}_{1}^{\dagger}\hat{a}_{1}^{\dagger}\hat{a}_{1}\hat{a}_{2}+\hat{a}_{1}^{\dagger}\hat{a}_{1}^{\dagger}\hat{a}_{2}\hat{a}_{1}\\
+\hat{a}_{2}^{\dagger}\hat{a}_{2}^{\dagger}\hat{a}_{1}\hat{a}_{2}+\hat{a}_{2}^{\dagger}\hat{a}_{2}^{\dagger}\hat{a}_{2}\hat{a}_{1}\\
+\hat{a}_{1}^{\dagger}\hat{a}_{2}^{\dagger}\hat{a}_{1}\hat{a}_{1}+\hat{a}_{2}^{\dagger}\hat{a}_{1}^{\dagger}\hat{a}_{1}\hat{a}_{1}\\
+\hat{a}_{1}^{\dagger}\hat{a}_{2}^{\dagger}\hat{a}_{2}\hat{a}_{2}+\hat{a}_{2}^{\dagger}\hat{a}_{1}^{\dagger}\hat{a}_{2}\hat{a}_{2}
\end{array}\right)\\
 & +\frac{g}{2}u_{N}\left(\begin{array}{c}
\hat{a}_{1}^{\dagger}\hat{a}_{2}^{\dagger}\hat{a}_{1}\hat{a}_{2}+\hat{a}_{1}^{\dagger}\hat{a}_{2}^{\dagger}\hat{a}_{2}\hat{a}_{1}\\
+\hat{a}_{2}^{\dagger}\hat{a}_{1}^{\dagger}\hat{a}_{1}\hat{a}_{2}+\hat{a}_{2}^{\dagger}\hat{a}_{1}^{\dagger}\hat{a}_{2}\hat{a}_{1}
\end{array}\right)\\
 & +\frac{g}{2}u_{P}\left(\hat{a}_{1}^{\dagger}\hat{a}_{1}^{\dagger}\hat{a}_{2}\hat{a}_{2}+\hat{a}_{2}^{\dagger}\hat{a}_{2}^{\dagger}\hat{a}_{1}\hat{a}_{1}\right)\\
 & +\frac{g}{2}u_{I}\left(\hat{a}_{1}^{\dagger}\hat{a}_{1}^{\dagger}\hat{a}_{1}\hat{a}_{1}+\hat{a}_{2}^{\dagger}\hat{a}_{2}^{\dagger}\hat{a}_{2}\hat{a}_{2}\right)\\
= & J\left(\hat{a}_{1}^{\dagger}\hat{a}_{2}+\hat{a}_{2}^{\dagger}\hat{a}_{1}\right)\\
 & +gu_{T}\left(\hat{a}_{1}^{\dagger}\hat{N}\hat{a}_{2}+\hat{a}_{2}^{\dagger}\hat{N}\hat{a}_{1}\right)\\
 & +\frac{g}{2}u_{N}\left(4\hat{N}_{1}\hat{N}_{2}\right)\\
 & +\frac{g}{2}u_{P}\left(\hat{a}_{1}^{\dagger}\hat{a}_{1}^{\dagger}\hat{a}_{2}\hat{a}_{2}+\hat{a}_{2}^{\dagger}\hat{a}_{2}^{\dagger}\hat{a}_{1}\hat{a}_{1}\right)\\
 & +\frac{g}{2}u_{I}\left(\hat{N}_{1}\left(\hat{N}_{1}-1\right)+\hat{N}_{2}\left(\hat{N}_{2}-1\right)\right),
\end{aligned}
\label{eq:ResultH0}
\end{equation}
which is the same as Eq. (\ref{eq:h0}).

\subsection{Derivation of Lipkin-Meshkov-Glick Hamiltonian - Equation (\ref{eq:LMKH})\label{subsec:Derivation3}}

Here, we recast $\hat{h}_{0}$ in terms of bosonic spin operators
defined as

\begin{equation}
\begin{array}{l}
\hat{S}_{x}=\left(\hat{a}_{2}^{\dagger}\hat{a}_{1}+\hat{a}_{1}^{\dagger}\hat{a}_{2}\right)/2\\
\hat{S}_{y}=\left(\hat{a}_{2}^{\dagger}\hat{a}_{1}-\hat{a}_{1}^{\dagger}\hat{a}_{2}\right)/2i\\
\hat{S}_{z}=\left(\hat{a}_{2}^{\dagger}\hat{a}_{2}-\hat{a}_{1}^{\dagger}\hat{a}_{1}\right)/2,
\end{array}
\end{equation}
which along with $\hat{N}=\hat{N}_{1}+\hat{N}_{2}$ and the standard
bosonic commutation rules for the mode annihilation, creation operators
enable the following substitutions to be made within $\hat{h}_{0}$
\begin{equation}
\begin{aligned}\hat{a}_{1}^{\dagger}\hat{a}_{2}+\hat{a}_{2}^{\dagger}\hat{a}_{1} & =2\hat{S}_{x}\\
\hat{N}_{1} & =\frac{1}{2}\hat{N}-\hat{S}_{z}\\
\hat{N}_{2} & =\frac{1}{2}\hat{N}+\hat{S}_{z}\\
\hat{a}_{1}^{\dagger}\hat{a}_{2} & =\hat{S}_{x}-i\hat{S}_{y}\\
\hat{a}_{2}^{\dagger}\hat{a}_{1} & =\hat{S}_{x}+i\hat{S}_{y}\\
\hat{a}_{1}^{\dagger}\hat{N}\hat{a}_{2} & =\hat{a}_{1}^{\dagger}\hat{a}_{2}(\hat{N}-1)\\
 & =\left(\hat{S}_{x}-i\hat{S}_{y}\right)(\hat{N}-1)\\
\hat{a}_{2}^{\dagger}\hat{N}\hat{a}_{1} & =\left(\hat{S}_{x}+i\hat{S}_{y}\right)(\hat{N}-1).
\end{aligned}
\end{equation}
From Eq. (\ref{eq:ResultH0}) we have 
\begin{equation}
\begin{aligned}\hat{h}_{0}= & J2\hat{S}_{x}\\
 & +gu_{T}\left(\left(\hat{S}_{x}-i\hat{S}_{y}\right)(\hat{N}-1)+\left(\hat{S}_{x}+i\hat{S}_{y}\right)(\hat{N}-1)\right)\\
 & +\frac{g}{2}u_{N}\left(4\left(\frac{1}{2}\hat{N}-\hat{S}_{z}\right)\left(\frac{1}{2}\hat{N}+\hat{S}_{z}\right)\right)\\
 & +\frac{g}{2}u_{P}\left(\left(\hat{S}_{x}-i\hat{S}_{y}\right)^{2}+\left(\hat{S}_{x}+i\hat{S}_{y}\right)^{2}\right)\\
 & +\frac{g}{2}u_{I}\left(\begin{array}{c}
\left(\frac{1}{2}\hat{N}-\hat{S}_{z}\right)\left(\frac{1}{2}\hat{N}-\hat{S}_{z}-1\right)\\
+\left(\frac{1}{2}\hat{N}+\hat{S}_{z}\right)\left(\frac{1}{2}\hat{N}+\hat{S}_{z}-1\right)
\end{array}\right)\\
= & J2\hat{S}_{x}\\
 & +gu_{T}2\hat{S}_{x}(\hat{N}-1)\\
 & +\frac{g}{2}u_{N}\left(4\left(\frac{1}{4}\hat{N}^{2}-\hat{S}_{z}^{2}\right)\right)\\
 & +\frac{g}{2}u_{P}2\left(\hat{S}_{x}^{2}-\hat{S}_{y}^{2}\right)\\
 & +\frac{g}{2}u_{I}2\left(\left(\frac{1}{2}\hat{N}\right)\left(\frac{1}{2}\hat{N}-1\right)+\hat{S}_{z}^{2}\right),
\end{aligned}
\end{equation}
so as we are only dealing with states which are eigenstates for the
total boson number, we can replace $\hat{N}$ by $N$.

After combining similar terms we get 
\begin{equation}
\begin{aligned}\hat{h}_{0}= & 2\left(J+gu_{T}(N-1)\right)\hat{S}_{x}\\
 & +g\left(-2u_{N}+u_{I}\right)\hat{S}_{z}^{2}\\
 & +gu_{P}\left(\hat{S}_{x}^{2}-\hat{S}_{y}^{2}\right)\\
 & +\frac{gN}{2}\left(u_{N}N+u_{I}\left(\frac{1}{2}N-1\right)\right),
\end{aligned}
\end{equation}
which can be written as 
\begin{equation}
\begin{aligned}\hat{h}_{0}= & \tilde{J}\left(-\hat{S}_{x}+\frac{\gamma}{N}\hat{S}_{z}^{2}\right)+E_{{\rm shift}}+\beta\left(\hat{S}_{x}^{2}-\hat{S}_{y}^{2}\right),\end{aligned}
\end{equation}
where 
\begin{equation}
\begin{aligned}\tilde{J} & =-2\left(J+gu_{T}(N-1)\right)\\
\gamma & =gN\left(-2u_{N}+u_{I}\right)/\tilde{J}\\
E_{{\rm shift}} & =\frac{gN}{2}\left(u_{N}N+u_{I}\left(\frac{1}{2}N-1\right)\right)\\
\beta & =gu_{P}.
\end{aligned}
\end{equation}
This is the same as Eq (\ref{eq:LMKH}) in the main body of the paper,
if the small term involving $\hat{S}_{x}^{2}-\hat{S}_{y}^{2}$ is
discarded.

\section{Many-Body Floquet mode solution\label{sec:MBF}}

We can define many-body Floquet states $\vec{\mathcal{F}}_{\nu}$
and Floquet energies $\mathcal{E}_{\nu}$ as the solutions of the
matrix equations (Eq. (\ref{eq:Full2mode})) 
\begin{equation}
\left(\underline{\mathcal{H}}-i\hbar\frac{\partial}{\partial t}\right)\vec{\mathcal{F}}_{\nu}=\mathcal{E}_{\nu}\vec{\mathcal{F}}_{\nu}.
\end{equation}
It can then be confirmed that a solution for the amplitudes $b_{n}(t)$
can be found in the form 
\begin{equation}
\vec{b}=\sum\mathcal{C}_{\nu}\exp\left(-i\mathcal{E}_{\nu}t/\hbar\right)\vec{\mathcal{F}}_{\nu},\label{eq:FloquetSoln}
\end{equation}
where the $\mathcal{C}_{\nu}$ are time-independent.

Substituting for $\vec{b}$ from Eq. (\ref{eq:FloquetSoln}) into
Eq. (\ref{eq:FinalAmpl}) gives 
\begin{equation}
\begin{aligned} & \left(\underline{\mathcal{H}}-i\hbar\frac{\partial}{\partial t}\right)\vec{b}\\
= & \sum_{\nu}\mathcal{C}_{\nu}\exp\left(-i\mathcal{E}_{\nu}t/\hbar\right)\left(\underline{\mathcal{H}}-i\hbar\frac{\partial}{\partial t}\right)\vec{\mathcal{F}}_{\nu}\\
 & -\sum_{\nu}\mathcal{C}_{\nu}\left(i\hbar\frac{\partial}{\partial t}\exp\left(-i\mathcal{E}_{\nu}t/\hbar\right)\right)\vec{\mathcal{F}}_{\nu}\\
= & \sum_{\nu}\mathcal{C}_{\nu}\exp\left(-i\mathcal{E}_{\nu}t/\hbar\right)\mathcal{E}_{\nu}\vec{\mathcal{F}}_{\nu}\\
 & -\sum_{\nu}\mathcal{C}_{\nu}\exp\left(-i\mathcal{E}_{\nu}t/\hbar\right)\mathcal{E}_{\nu}\vec{\mathcal{F}}_{\nu}\\
= & 0
\end{aligned}
\end{equation}
as required. The initial condition gives $\ensuremath{\mathcal{C}_{\nu}=\vec{\mathcal{F}}{}_{\nu}^{\dagger}\cdot\vec{b}(0)}$.

\section{Quantum revival\label{sec:revival}}

\begin{figure}
\includegraphics[width=0.98\columnwidth]{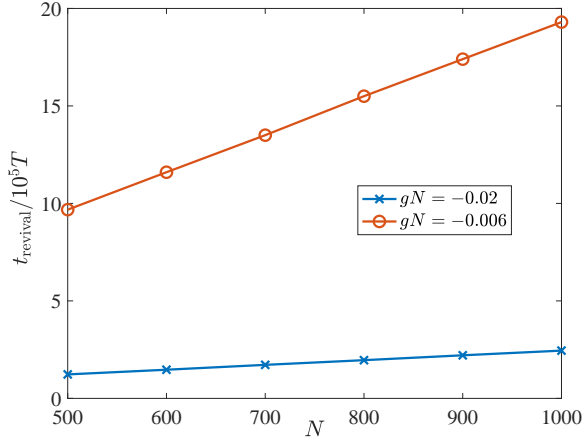}\caption{The revival time as a function of particle number $N$. The blue crosses
(red circles) show the revival time for $gN=-0.02$ ($-0.006$), respectively.
The initial state is given by $|0,N\rangle$.\label{fig:Revival}}
\end{figure}

The quantum revival time can be obtained from the spectrum \citep{Zakrzewski2002PR,Dalton2012JMO}
\begin{equation}
T_{\text{revival }}=\frac{2\pi\hbar}{\left|\left(\mathrm{d}^{2}\mathcal{E}_{\nu}/\mathrm{d}\nu^{2}\right)\left(\nu_{0}\right)\right|},
\end{equation}
where $\mathcal{E_{\nu}}$ are understood to be taken from the same
branch (so that there is no near-degeneracy), and $\nu$ is the index
for ascending eigenenergies. $\nu_{0}$ indicates the closest $\mathcal{E}_{\nu_{0}}$
to the initial energy $E_{{\rm ini}}$, and the second derivative
can be approximated by $d^{2}\mathcal{E}_{\nu}/d\nu^{2}\approx\mathcal{E}_{\nu-1}+\mathcal{E}_{\nu+1}-2\mathcal{E}_{\nu}$.
Interestingly, we numerically find that the revival time linearly
depends on the particle number $N$, as shown in Fig. \ref{fig:Revival}.
In other words, in the infinite $N$ limit, the revival would not
be accessible.

\section{Initial product states\label{sec:mf}}

\begin{figure}
\includegraphics[width=0.98\columnwidth]{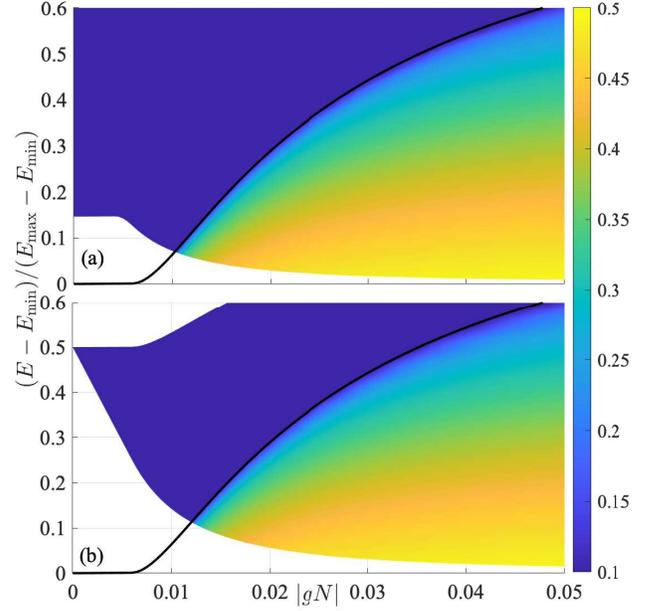}\caption{Subharmonic response $|m_{G}(\omega/2)|$ (colour bar) as a function
of normalized $E_{{\rm ini}}$ and $gN$. The black curve shows the
normalized $\mathcal{E}_{{\rm edge}}$ as a function of $|gN|$, which
coincides with the phase boundary for both initial product states
$|\{\theta,\varphi=0.25\pi\}\rangle$ in (a) and $|\{\theta,\varphi=0.5\pi\}\rangle$
in (b).\label{fig:PDdiffvarphi}}
\end{figure}

For initial states $|\Theta(0)\rangle=|\{\theta,\varphi\}\rangle$
with $\theta\in[0,\pi)$ and $\varphi\in[0,\pi)$ given by Eq. (\ref{eq:theta_state}),
all atoms occupy the same mode $\Lambda_{2}$ $=$$\sin\theta e^{i\varphi}\Phi_{1}+\cos\theta\Phi_{2}$,
and hence the energy can be given by a mean-field approach in the
large $N$ limit, i.e., by replacing the creation and annhilation
operators in the Hamiltonian (\ref{eq:h0}) via

\begin{equation}
\hat{a}_{1},\hat{a}_{1}^{\dagger}\rightarrow\sqrt{N}\sin\theta e^{\pm i\varphi},\hat{a}_{2},\hat{a}_{2}^{\dagger}\rightarrow\sqrt{N}\cos\theta.
\end{equation}
Neglecting the small terms associated with $u_{P}$, the energy functional
is given by

\begin{equation}
\frac{E_{{\rm ini}}(\theta,\varphi)-E_{{\rm shift}}}{N}=-\frac{\tilde{J}}{2}\sin2\theta\cos\varphi+\frac{\tilde{\gamma}}{4}\cos^{2}2\theta,
\end{equation}
where $\tilde{\gamma}=gN(u_{I}-2u_{N})$. Inserting the definitions
of $\tilde{J}$ and $E_{{\rm shift}}$ given in the main text 
\begin{equation}
\tilde{J}=-2\left[J+gu_{T}(N-1)\right]\approx-2[J+gNu_{T}]
\end{equation}
and 
\begin{equation}
\frac{E_{{\rm shift}}}{N}=\frac{1}{2}gN\left(\frac{u_{I}}{2}-\frac{u_{I}}{N}+u_{N}\right)\approx\frac{1}{2}gN\left(\frac{u_{I}}{2}+u_{N}\right)
\end{equation}
leads to 
\begin{equation}
\begin{aligned}\frac{E_{{\rm ini}}(\theta,\varphi)}{N} & \approx J\sin2\theta\cos(\varphi)+gN\left[\frac{u_{I}}{2}+\right.\\
 & \left.u_{T}\sin2\theta\cos(\varphi)+\frac{(2u_{N}-u_{I})}{4}\sin^{2}2\theta\right],
\end{aligned}
\label{eq:EiniAppendix}
\end{equation}
which agrees with Eq. (\ref{eq:Eini}) in the main text. Equating
$E_{{\rm ini}}(\theta,\varphi)$ {[}given in Eq. (\ref{eq:Eini})
or Eq. (\ref{eq:EiniAppendix}){]} with the broken symmetry edge $\mathcal{E}_{{\rm edge}}$
{[}given in Eq. (\ref{eq:Eedge}){]} leads to the critical interaction
strength as a function of $\theta$ for a given $\varphi$ 
\begin{equation}
g_{c}(\theta;\varphi)N=-\frac{J(1+\sin2\theta\cos\varphi)}{(u_{I}/4-u_{N}/2)\cos^{2}2\theta+u_{T}(1+\sin2\theta\cos\varphi)},\label{eq:gcN}
\end{equation}
which is shown in Fig. \ref{fig:gcN}. It is also important to note
that the phase diagram does not depend explicitly on $\{\theta,\varphi\}$,
where the subharmonic response $|m_{G}(\omega/2)|$ is essentially
zero (nonzero) for $E_{{\rm ini}}(\theta,\varphi)>E_{{\rm {\rm edge}}}$
($E_{{\rm ini}}(\theta,\varphi)\le E_{{\rm {\rm edge}}}$) for a given
$|gN|.$ Figures \ref{fig:phasediagram} (b), \ref{fig:PDdiffvarphi}
(a) and (b) show $|m_{G}(\omega/2)|$ for different initial states
$|\{\theta,\varphi=0\}\rangle$, $|\{\theta,\varphi=0.25\pi\}\rangle$
and $|\{\theta,\varphi=0.5\pi\}\rangle$, respectively, which agree
with each other. However, the condition $\varphi\ne0$ would limit
the range of $E_{{\rm ini}}(\theta,\varphi)$ that can be accessed,
which is indicated by the white areas (no data) in Figs. \ref{fig:PDdiffvarphi}
(a) and (b).

In the large $N$ limit, since the spectrum is bounded from both below
and above, the minimum and maximum energy state should also be well
approximated by the mean-field expression, i.e., $E_{{\rm max}}=\max\left[E_{{\rm ini}}(\theta,\varphi)\right]$
and $E_{{\rm min}}=\min\left[E_{{\rm ini}}(\theta,\varphi)\right]$.
These expressions are explicitly given by

\begin{equation}
E_{{\rm max}}\approx E_{{\rm shift}}+\frac{|\tilde{J}|N}{2},
\end{equation}
and for the negative $gN$ considered in this work

\begin{equation}
E_{{\rm min}}\approx\begin{cases}
E_{{\rm shift}}-\frac{|\tilde{J}|N}{2}, & |\tilde{J}|\ge-\tilde{\gamma}\\
E_{{\rm shift}}+N\left(\frac{\tilde{\gamma}}{4}+\frac{|\tilde{J}|^{2}}{4\tilde{\gamma}}\right), & |\tilde{J}|<-\tilde{\gamma}
\end{cases},
\end{equation}
whose expression in the second line can also be explicitly written
as $gN^{2}u_{I}/2+\tilde{J}^{2}N/gN(u_{I}-2u_{N})$. In this work,
we focus on the case $J+gNu_{T}>0$, therefore; the critical condition
gives 
\begin{equation}
g_{b}N=-2J/(u_{I}-2u_{N}+2u_{T}).\label{eq:gbN}
\end{equation}
One can notice that the initial state corresponding to $E_{{\rm min}}$
actually corresponds to the ``Wannier initial condition'' in our
previous study \citep{WangNJP2021}. We also find here that the minimum
value of $|g_{c}N|$ for DTC formation based on Wannier initial conditions
is given by $|g_{c}N|=|g_{b}N|\approx0.006$. This corresponds to
$E_{{\rm edge}}=E_{{\rm min}}$ in Fig. \ref{fig:phasediagram} (b).

\section{Finite size effect\label{sec:FiniteSize}}

\begin{figure}
\includegraphics[width=0.98\columnwidth]{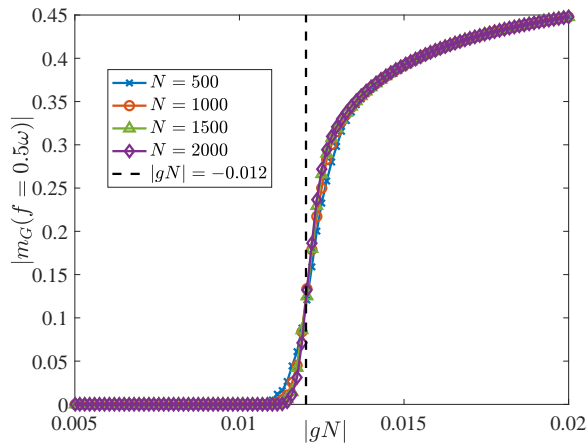}\caption{Steady states' subharmonic response signal as a function of $|gN|$
for initial state $|0,N\rangle$ for different $N$ denoted in the
figure. The black dashed vertical line indicates the critical interaction
strength $|gN|=0.012$. \label{fig:FiniteSize}}
\end{figure}

In Ref. \citep{Pizz2020iPRB}, it is noticed that the amplitude of
the subharmonic response of a DTC in a many-body quantum scar system
is exponentially suppressed with system size, and in principle would
disappear in the thermodynamic limit. In contrast, the subharmonic
response of a DTC in a MBL system does not decay with respect to system
size. It is thus interesting to study the finite size effect of our
system, whose eigenspectra are different from both many-body quantum
scar and MBL systems. Figure \ref{fig:FiniteSize} shows the subharmonic
response as a function of $|gN|$ for different $N$. We observe that
the subharmonic response in the symmetry-broken phase does not decay
with respect to $N$, similar to the DTC in MBL systems.

\bibliography{RefsTC2Mode}

\begin{thebibliography}{57}%
\makeatletter
\providecommand \@ifxundefined [1]{%
 \@ifx{#1\undefined}
}%
\providecommand \@ifnum [1]{%
 \ifnum #1\expandafter \@firstoftwo
 \else \expandafter \@secondoftwo
 \fi
}%
\providecommand \@ifx [1]{%
 \ifx #1\expandafter \@firstoftwo
 \else \expandafter \@secondoftwo
 \fi
}%
\providecommand \natexlab [1]{#1}%
\providecommand \enquote  [1]{``#1''}%
\providecommand \bibnamefont  [1]{#1}%
\providecommand \bibfnamefont [1]{#1}%
\providecommand \citenamefont [1]{#1}%
\providecommand \href@noop [0]{\@secondoftwo}%
\providecommand \href [0]{\begingroup \@sanitize@url \@href}%
\providecommand \@href[1]{\@@startlink{#1}\@@href}%
\providecommand \@@href[1]{\endgroup#1\@@endlink}%
\providecommand \@sanitize@url [0]{\catcode `\\12\catcode `\$12\catcode
  `\&12\catcode `\#12\catcode `\^12\catcode `\_12\catcode `\%12\relax}%
\providecommand \@@startlink[1]{}%
\providecommand \@@endlink[0]{}%
\providecommand \url  [0]{\begingroup\@sanitize@url \@url }%
\providecommand \@url [1]{\endgroup\@href {#1}{\urlprefix }}%
\providecommand \urlprefix  [0]{URL }%
\providecommand \Eprint [0]{\href }%
\providecommand \doibase [0]{http://dx.doi.org/}%
\providecommand \selectlanguage [0]{\@gobble}%
\providecommand \bibinfo  [0]{\@secondoftwo}%
\providecommand \bibfield  [0]{\@secondoftwo}%
\providecommand \translation [1]{[#1]}%
\providecommand \BibitemOpen [0]{}%
\providecommand \bibitemStop [0]{}%
\providecommand \bibitemNoStop [0]{.\EOS\space}%
\providecommand \EOS [0]{\spacefactor3000\relax}%
\providecommand \BibitemShut  [1]{\csname bibitem#1\endcsname}%
\let\auto@bib@innerbib\@empty
\bibitem [{\citenamefont {Wilczek}(2012)}]{Wilczek2012PRL}%
  \BibitemOpen
  \bibfield  {author} {\bibinfo {author} {\bibfnamefont {Frank}\ \bibnamefont
  {Wilczek}},\ }\bibfield  {title} {\enquote {\bibinfo {title} {Quantum time
  crystals},}\ }\href@noop {} {\bibfield  {journal} {\bibinfo  {journal} {Phys.
  Rev. Lett.}\ }\textbf {\bibinfo {volume} {109}},\ \bibinfo {pages} {160401}
  (\bibinfo {year} {2012})}\BibitemShut {NoStop}%
\bibitem [{\citenamefont {Watanabe}\ and\ \citenamefont
  {Oshikawa}(2015)}]{WatanabePRL2015}%
  \BibitemOpen
  \bibfield  {author} {\bibinfo {author} {\bibfnamefont {Haruki}\ \bibnamefont
  {Watanabe}}\ and\ \bibinfo {author} {\bibfnamefont {Masaki}\ \bibnamefont
  {Oshikawa}},\ }\bibfield  {title} {\enquote {\bibinfo {title} {Absence of
  quantum time crystals},}\ }\href@noop {} {\bibfield  {journal} {\bibinfo
  {journal} {Phys. Rev. Lett.}\ }\textbf {\bibinfo {volume} {114}},\ \bibinfo
  {pages} {251603} (\bibinfo {year} {2015})}\BibitemShut {NoStop}%
\bibitem [{\citenamefont {Kozin}\ and\ \citenamefont
  {Kyriienko}(2019)}]{Kozin2019PRL}%
  \BibitemOpen
  \bibfield  {author} {\bibinfo {author} {\bibfnamefont {Valerii~K.}\
  \bibnamefont {Kozin}}\ and\ \bibinfo {author} {\bibfnamefont {Oleksandr}\
  \bibnamefont {Kyriienko}},\ }\bibfield  {title} {\enquote {\bibinfo {title}
  {Quantum time crystals from {Hamiltonians} with long-range interactions},}\
  }\href@noop {} {\bibfield  {journal} {\bibinfo  {journal} {Phys. Rev. Lett.}\
  }\textbf {\bibinfo {volume} {123}},\ \bibinfo {pages} {210602} (\bibinfo
  {year} {2019})}\BibitemShut {NoStop}%
\bibitem [{\citenamefont {Sacha}(2015)}]{Sacha2015PRA}%
  \BibitemOpen
  \bibfield  {author} {\bibinfo {author} {\bibfnamefont {Krzysztof}\
  \bibnamefont {Sacha}},\ }\bibfield  {title} {\enquote {\bibinfo {title}
  {Modeling spontaneous breaking of time-translation symmetry},}\ }\href@noop
  {} {\bibfield  {journal} {\bibinfo  {journal} {Phys. Rev. A}\ }\textbf
  {\bibinfo {volume} {91}},\ \bibinfo {pages} {033617} (\bibinfo {year}
  {2015})}\BibitemShut {NoStop}%
\bibitem [{\citenamefont {Khemani}\ \emph {et~al.}(2016)\citenamefont
  {Khemani}, \citenamefont {Lazarides}, \citenamefont {Moessner},\ and\
  \citenamefont {Sondhi}}]{Khemani2016PRL}%
  \BibitemOpen
  \bibfield  {author} {\bibinfo {author} {\bibfnamefont {Vedika}\ \bibnamefont
  {Khemani}}, \bibinfo {author} {\bibfnamefont {Achilleas}\ \bibnamefont
  {Lazarides}}, \bibinfo {author} {\bibfnamefont {Roderich}\ \bibnamefont
  {Moessner}}, \ and\ \bibinfo {author} {\bibfnamefont {S.~L.}\ \bibnamefont
  {Sondhi}},\ }\bibfield  {title} {\enquote {\bibinfo {title} {Phase structure
  of driven quantum systems},}\ }\href@noop {} {\bibfield  {journal} {\bibinfo
  {journal} {Phys. Rev. Lett.}\ }\textbf {\bibinfo {volume} {116}},\ \bibinfo
  {pages} {250401} (\bibinfo {year} {2016})}\BibitemShut {NoStop}%
\bibitem [{\citenamefont {Else}\ \emph {et~al.}(2016)\citenamefont {Else},
  \citenamefont {Bauer},\ and\ \citenamefont {Nayak}}]{Else2016PRL}%
  \BibitemOpen
  \bibfield  {author} {\bibinfo {author} {\bibfnamefont {Dominic~V.}\
  \bibnamefont {Else}}, \bibinfo {author} {\bibfnamefont {Bela}\ \bibnamefont
  {Bauer}}, \ and\ \bibinfo {author} {\bibfnamefont {Chetan}\ \bibnamefont
  {Nayak}},\ }\bibfield  {title} {\enquote {\bibinfo {title} {Floquet time
  crystals},}\ }\href@noop {} {\bibfield  {journal} {\bibinfo  {journal} {Phys.
  Rev. Lett.}\ }\textbf {\bibinfo {volume} {117}},\ \bibinfo {pages} {090402}
  (\bibinfo {year} {2016})}\BibitemShut {NoStop}%
\bibitem [{\citenamefont {Yao}\ \emph {et~al.}(2017)\citenamefont {Yao},
  \citenamefont {Potter}, \citenamefont {Potirniche},\ and\ \citenamefont
  {Vishwanath}}]{Yao2017PRL}%
  \BibitemOpen
  \bibfield  {author} {\bibinfo {author} {\bibfnamefont {N.~Y.}\ \bibnamefont
  {Yao}}, \bibinfo {author} {\bibfnamefont {A.~C.}\ \bibnamefont {Potter}},
  \bibinfo {author} {\bibfnamefont {I.-D.}\ \bibnamefont {Potirniche}}, \ and\
  \bibinfo {author} {\bibfnamefont {A.}~\bibnamefont {Vishwanath}},\ }\bibfield
   {title} {\enquote {\bibinfo {title} {Discrete time crystals: Rigidity,
  criticality, and realizations},}\ }\href@noop {} {\bibfield  {journal}
  {\bibinfo  {journal} {Phys. Rev. Lett.}\ }\textbf {\bibinfo {volume} {118}},\
  \bibinfo {pages} {030401} (\bibinfo {year} {2017})}\BibitemShut {NoStop}%
\bibitem [{\citenamefont {Zhang}\ \emph {et~al.}(2017)\citenamefont {Zhang},
  \citenamefont {Hess}, \citenamefont {Kyprianidis}, \citenamefont {Becker},
  \citenamefont {Lee}, \citenamefont {Smith}, \citenamefont {Pagano},
  \citenamefont {Potirniche}, \citenamefont {Potter}, \citenamefont
  {Vishwanath}, \citenamefont {Yao},\ and\ \citenamefont
  {Monroe}}]{Monroe2017Nature}%
  \BibitemOpen
  \bibfield  {author} {\bibinfo {author} {\bibfnamefont {J.}~\bibnamefont
  {Zhang}}, \bibinfo {author} {\bibfnamefont {P.~W.}\ \bibnamefont {Hess}},
  \bibinfo {author} {\bibfnamefont {A.}~\bibnamefont {Kyprianidis}}, \bibinfo
  {author} {\bibfnamefont {P.}~\bibnamefont {Becker}}, \bibinfo {author}
  {\bibfnamefont {A.}~\bibnamefont {Lee}}, \bibinfo {author} {\bibfnamefont
  {J.}~\bibnamefont {Smith}}, \bibinfo {author} {\bibfnamefont
  {G.}~\bibnamefont {Pagano}}, \bibinfo {author} {\bibfnamefont {I.-D.}\
  \bibnamefont {Potirniche}}, \bibinfo {author} {\bibfnamefont {A.~C.}\
  \bibnamefont {Potter}}, \bibinfo {author} {\bibfnamefont {A.}~\bibnamefont
  {Vishwanath}}, \bibinfo {author} {\bibfnamefont {N.~Y.}\ \bibnamefont {Yao}},
  \ and\ \bibinfo {author} {\bibfnamefont {C.}~\bibnamefont {Monroe}},\
  }\bibfield  {title} {\enquote {\bibinfo {title} {Observation of a discrete
  time crystal},}\ }\href@noop {} {\bibfield  {journal} {\bibinfo  {journal}
  {Nature}\ }\textbf {\bibinfo {volume} {543}},\ \bibinfo {pages} {217}
  (\bibinfo {year} {2017})}\BibitemShut {NoStop}%
\bibitem [{\citenamefont {Choi}\ \emph {et~al.}(2017)\citenamefont {Choi},
  \citenamefont {Choi}, \citenamefont {Landig}, \citenamefont {Kucsko},
  \citenamefont {Zhou}, \citenamefont {Isoya}, \citenamefont {Jelezko},
  \citenamefont {Onoda}, \citenamefont {Sumiya}, \citenamefont {Khemani},
  \citenamefont {von Keyserlingk}, \citenamefont {Yao}, \citenamefont
  {Demler},\ and\ \citenamefont {Lukin}}]{Lukin2017Nature}%
  \BibitemOpen
  \bibfield  {author} {\bibinfo {author} {\bibfnamefont {Soonwon}\ \bibnamefont
  {Choi}}, \bibinfo {author} {\bibfnamefont {Joonhee}\ \bibnamefont {Choi}},
  \bibinfo {author} {\bibfnamefont {Renate}\ \bibnamefont {Landig}}, \bibinfo
  {author} {\bibfnamefont {Georg}\ \bibnamefont {Kucsko}}, \bibinfo {author}
  {\bibfnamefont {Hengyun}\ \bibnamefont {Zhou}}, \bibinfo {author}
  {\bibfnamefont {Junichi}\ \bibnamefont {Isoya}}, \bibinfo {author}
  {\bibfnamefont {Fedor}\ \bibnamefont {Jelezko}}, \bibinfo {author}
  {\bibfnamefont {Shinobu}\ \bibnamefont {Onoda}}, \bibinfo {author}
  {\bibfnamefont {Hitoshi}\ \bibnamefont {Sumiya}}, \bibinfo {author}
  {\bibfnamefont {Vedika}\ \bibnamefont {Khemani}}, \bibinfo {author}
  {\bibfnamefont {Curt}\ \bibnamefont {von Keyserlingk}}, \bibinfo {author}
  {\bibfnamefont {Norman~Y.}\ \bibnamefont {Yao}}, \bibinfo {author}
  {\bibfnamefont {Eugene}\ \bibnamefont {Demler}}, \ and\ \bibinfo {author}
  {\bibfnamefont {Mikhail~D.}\ \bibnamefont {Lukin}},\ }\bibfield  {title}
  {\enquote {\bibinfo {title} {Observation of a discrete time crystal},}\
  }\href@noop {} {\bibfield  {journal} {\bibinfo  {journal} {Nature}\ }\textbf
  {\bibinfo {volume} {543}},\ \bibinfo {pages} {221} (\bibinfo {year}
  {2017})}\BibitemShut {NoStop}%
\bibitem [{\citenamefont {Rovny}\ \emph
  {et~al.}(2018{\natexlab{a}})\citenamefont {Rovny}, \citenamefont {Blum},\
  and\ \citenamefont {Barrett}}]{Barrett2018PRL}%
  \BibitemOpen
  \bibfield  {author} {\bibinfo {author} {\bibfnamefont {Jared}\ \bibnamefont
  {Rovny}}, \bibinfo {author} {\bibfnamefont {Robert~L.}\ \bibnamefont {Blum}},
  \ and\ \bibinfo {author} {\bibfnamefont {Sean~E.}\ \bibnamefont {Barrett}},\
  }\bibfield  {title} {\enquote {\bibinfo {title} {Observation of
  discrete-time-crystal signatures in an ordered dipolar many-body system},}\
  }\href@noop {} {\bibfield  {journal} {\bibinfo  {journal} {Phys. Rev. Lett.}\
  }\textbf {\bibinfo {volume} {120}},\ \bibinfo {pages} {180603} (\bibinfo
  {year} {2018}{\natexlab{a}})}\BibitemShut {NoStop}%
\bibitem [{\citenamefont {Pal}\ \emph {et~al.}(2018)\citenamefont {Pal},
  \citenamefont {Nishad}, \citenamefont {Mahesh},\ and\ \citenamefont
  {Sreejith}}]{Sreejith2018PRL}%
  \BibitemOpen
  \bibfield  {author} {\bibinfo {author} {\bibfnamefont {Soham}\ \bibnamefont
  {Pal}}, \bibinfo {author} {\bibfnamefont {Naveen}\ \bibnamefont {Nishad}},
  \bibinfo {author} {\bibfnamefont {T.~S.}\ \bibnamefont {Mahesh}}, \ and\
  \bibinfo {author} {\bibfnamefont {G.~J.}\ \bibnamefont {Sreejith}},\
  }\bibfield  {title} {\enquote {\bibinfo {title} {Temporal order in
  periodically driven spins in star-shaped clusters},}\ }\href@noop {}
  {\bibfield  {journal} {\bibinfo  {journal} {Phys. Rev. Lett.}\ }\textbf
  {\bibinfo {volume} {120}},\ \bibinfo {pages} {180602} (\bibinfo {year}
  {2018})}\BibitemShut {NoStop}%
\bibitem [{\citenamefont {Rovny}\ \emph
  {et~al.}(2018{\natexlab{b}})\citenamefont {Rovny}, \citenamefont {Blum},\
  and\ \citenamefont {Barrett}}]{Barrett2018PRB}%
  \BibitemOpen
  \bibfield  {author} {\bibinfo {author} {\bibfnamefont {Jared}\ \bibnamefont
  {Rovny}}, \bibinfo {author} {\bibfnamefont {Robert~L.}\ \bibnamefont {Blum}},
  \ and\ \bibinfo {author} {\bibfnamefont {Sean~E.}\ \bibnamefont {Barrett}},\
  }\bibfield  {title} {\enquote {\bibinfo {title} {$^{31}\mathrm{P}$ {NMR}
  study of discrete time-crystalline signatures in an ordered crystal of
  ammonium dihydrogen phosphate},}\ }\href@noop {} {\bibfield  {journal}
  {\bibinfo  {journal} {Phys. Rev. B}\ }\textbf {\bibinfo {volume} {97}},\
  \bibinfo {pages} {184301} (\bibinfo {year} {2018}{\natexlab{b}})}\BibitemShut
  {NoStop}%
\bibitem [{\citenamefont {Smits}\ \emph {et~al.}(2018)\citenamefont {Smits},
  \citenamefont {Liao}, \citenamefont {Stoof},\ and\ \citenamefont {van~der
  Straten}}]{Stoof2018PRL}%
  \BibitemOpen
  \bibfield  {author} {\bibinfo {author} {\bibfnamefont {J.}~\bibnamefont
  {Smits}}, \bibinfo {author} {\bibfnamefont {L.}~\bibnamefont {Liao}},
  \bibinfo {author} {\bibfnamefont {H.~T.~C.}\ \bibnamefont {Stoof}}, \ and\
  \bibinfo {author} {\bibfnamefont {P.}~\bibnamefont {van~der Straten}},\
  }\bibfield  {title} {\enquote {\bibinfo {title} {Observation of a space-time
  crystal in a superfluid quantum gas},}\ }\href@noop {} {\bibfield  {journal}
  {\bibinfo  {journal} {Phys. Rev. Lett.}\ }\textbf {\bibinfo {volume} {121}},\
  \bibinfo {pages} {185301} (\bibinfo {year} {2018})}\BibitemShut {NoStop}%
\bibitem [{\citenamefont {Liao}\ \emph {et~al.}(2019)\citenamefont {Liao},
  \citenamefont {Smits}, \citenamefont {van~der Straten},\ and\ \citenamefont
  {Stoof}}]{Stoof2019PRA}%
  \BibitemOpen
  \bibfield  {author} {\bibinfo {author} {\bibfnamefont {L.}~\bibnamefont
  {Liao}}, \bibinfo {author} {\bibfnamefont {J.}~\bibnamefont {Smits}},
  \bibinfo {author} {\bibfnamefont {P.}~\bibnamefont {van~der Straten}}, \ and\
  \bibinfo {author} {\bibfnamefont {H.~T.~C.}\ \bibnamefont {Stoof}},\
  }\bibfield  {title} {\enquote {\bibinfo {title} {Dynamics of a space-time
  crystal in an atomic bose-einstein condensate},}\ }\href@noop {} {\bibfield
  {journal} {\bibinfo  {journal} {Phys. Rev. A}\ }\textbf {\bibinfo {volume}
  {99}},\ \bibinfo {pages} {013625} (\bibinfo {year} {2019})}\BibitemShut
  {NoStop}%
\bibitem [{\citenamefont {Smits}\ \emph {et~al.}(2020)\citenamefont {Smits},
  \citenamefont {Stoof},\ and\ \citenamefont {van~der Straten}}]{Stoof2020NJP}%
  \BibitemOpen
  \bibfield  {author} {\bibinfo {author} {\bibfnamefont {J.}~\bibnamefont
  {Smits}}, \bibinfo {author} {\bibfnamefont {H.~T.~C.}\ \bibnamefont {Stoof}},
  \ and\ \bibinfo {author} {\bibfnamefont {P.}~\bibnamefont {van~der
  Straten}},\ }\bibfield  {title} {\enquote {\bibinfo {title} {On the long-term
  stability of space-time crystals},}\ }\href@noop {} {\bibfield  {journal}
  {\bibinfo  {journal} {New J. Phys.}\ }\textbf {\bibinfo {volume} {22}},\
  \bibinfo {pages} {105001} (\bibinfo {year} {2020})}\BibitemShut {NoStop}%
\bibitem [{\citenamefont {Else}\ \emph {et~al.}(2017)\citenamefont {Else},
  \citenamefont {Bauer},\ and\ \citenamefont {Nayak}}]{Else2017PRX}%
  \BibitemOpen
  \bibfield  {author} {\bibinfo {author} {\bibfnamefont {Dominic~V.}\
  \bibnamefont {Else}}, \bibinfo {author} {\bibfnamefont {Bela}\ \bibnamefont
  {Bauer}}, \ and\ \bibinfo {author} {\bibfnamefont {Chetan}\ \bibnamefont
  {Nayak}},\ }\bibfield  {title} {\enquote {\bibinfo {title} {Prethermal phases
  of matter protected by time-translation symmetry},}\ }\href@noop {}
  {\bibfield  {journal} {\bibinfo  {journal} {Phys. Rev. X}\ }\textbf {\bibinfo
  {volume} {7}},\ \bibinfo {pages} {011026} (\bibinfo {year}
  {2017})}\BibitemShut {NoStop}%
\bibitem [{\citenamefont {Russomanno}\ \emph {et~al.}(2017)\citenamefont
  {Russomanno}, \citenamefont {Iemini}, \citenamefont {Dalmonte},\ and\
  \citenamefont {Fazio}}]{Fazio2017PRB}%
  \BibitemOpen
  \bibfield  {author} {\bibinfo {author} {\bibfnamefont {Angelo}\ \bibnamefont
  {Russomanno}}, \bibinfo {author} {\bibfnamefont {Fernando}\ \bibnamefont
  {Iemini}}, \bibinfo {author} {\bibfnamefont {Marcello}\ \bibnamefont
  {Dalmonte}}, \ and\ \bibinfo {author} {\bibfnamefont {Rosario}\ \bibnamefont
  {Fazio}},\ }\bibfield  {title} {\enquote {\bibinfo {title} {Floquet time
  crystal in the lipkin-meshkov-glick model},}\ }\href@noop {} {\bibfield
  {journal} {\bibinfo  {journal} {Phys. Rev. B}\ }\textbf {\bibinfo {volume}
  {95}},\ \bibinfo {pages} {214307} (\bibinfo {year} {2017})}\BibitemShut
  {NoStop}%
\bibitem [{\citenamefont {Matus}\ and\ \citenamefont
  {Sacha}(2019)}]{Sacha2019PRA}%
  \BibitemOpen
  \bibfield  {author} {\bibinfo {author} {\bibfnamefont {Pawe\l{}}\
  \bibnamefont {Matus}}\ and\ \bibinfo {author} {\bibfnamefont {Krzysztof}\
  \bibnamefont {Sacha}},\ }\bibfield  {title} {\enquote {\bibinfo {title}
  {Fractional time crystals},}\ }\href@noop {} {\bibfield  {journal} {\bibinfo
  {journal} {Phys. Rev. A}\ }\textbf {\bibinfo {volume} {99}},\ \bibinfo
  {pages} {033626} (\bibinfo {year} {2019})}\BibitemShut {NoStop}%
\bibitem [{\citenamefont {Zhu}\ \emph {et~al.}(2019)\citenamefont {Zhu},
  \citenamefont {Marino}, \citenamefont {Yao}, \citenamefont {Lukin},\ and\
  \citenamefont {Demler}}]{Demler2019NJP}%
  \BibitemOpen
  \bibfield  {author} {\bibinfo {author} {\bibfnamefont {Bihui}\ \bibnamefont
  {Zhu}}, \bibinfo {author} {\bibfnamefont {Jamir}\ \bibnamefont {Marino}},
  \bibinfo {author} {\bibfnamefont {Norman~Y}\ \bibnamefont {Yao}}, \bibinfo
  {author} {\bibfnamefont {Mikhail~D}\ \bibnamefont {Lukin}}, \ and\ \bibinfo
  {author} {\bibfnamefont {Eugene~A}\ \bibnamefont {Demler}},\ }\bibfield
  {title} {\enquote {\bibinfo {title} {Dicke time crystals in
  driven-dissipative quantum many-body systems},}\ }\href@noop {} {\bibfield
  {journal} {\bibinfo  {journal} {New J. Phys.}\ }\textbf {\bibinfo {volume}
  {21}},\ \bibinfo {pages} {073028} (\bibinfo {year} {2019})}\BibitemShut
  {NoStop}%
\bibitem [{\citenamefont {Sun}\ \emph {et~al.}(2019)\citenamefont {Sun},
  \citenamefont {He}, \citenamefont {Gong}, \citenamefont {Teh}, \citenamefont
  {Reid},\ and\ \citenamefont {Drummond}}]{Drummond2019NJP}%
  \BibitemOpen
  \bibfield  {author} {\bibinfo {author} {\bibfnamefont {Feng-Xiao}\
  \bibnamefont {Sun}}, \bibinfo {author} {\bibfnamefont {Qiongyi}\ \bibnamefont
  {He}}, \bibinfo {author} {\bibfnamefont {Qihuang}\ \bibnamefont {Gong}},
  \bibinfo {author} {\bibfnamefont {Run~Yan}\ \bibnamefont {Teh}}, \bibinfo
  {author} {\bibfnamefont {Margaret~D}\ \bibnamefont {Reid}}, \ and\ \bibinfo
  {author} {\bibfnamefont {Peter~D}\ \bibnamefont {Drummond}},\ }\bibfield
  {title} {\enquote {\bibinfo {title} {Discrete time symmetry breaking in
  quantum circuits: exact solutions and tunneling},}\ }\href@noop {} {\bibfield
   {journal} {\bibinfo  {journal} {New J. Phys.}\ }\textbf {\bibinfo {volume}
  {21}},\ \bibinfo {pages} {093035} (\bibinfo {year} {2019})}\BibitemShut
  {NoStop}%
\bibitem [{\citenamefont {Machado}\ \emph {et~al.}(2020)\citenamefont
  {Machado}, \citenamefont {Else}, \citenamefont {Kahanamoku-Meyer},
  \citenamefont {Nayak},\ and\ \citenamefont {Yao}}]{Else2020PRX}%
  \BibitemOpen
  \bibfield  {author} {\bibinfo {author} {\bibfnamefont {Francisco}\
  \bibnamefont {Machado}}, \bibinfo {author} {\bibfnamefont {Dominic~V.}\
  \bibnamefont {Else}}, \bibinfo {author} {\bibfnamefont {Gregory~D.}\
  \bibnamefont {Kahanamoku-Meyer}}, \bibinfo {author} {\bibfnamefont {Chetan}\
  \bibnamefont {Nayak}}, \ and\ \bibinfo {author} {\bibfnamefont {Norman~Y.}\
  \bibnamefont {Yao}},\ }\bibfield  {title} {\enquote {\bibinfo {title}
  {Long-range prethermal phases of nonequilibrium matter},}\ }\href@noop {}
  {\bibfield  {journal} {\bibinfo  {journal} {Phys. Rev. X}\ }\textbf {\bibinfo
  {volume} {10}},\ \bibinfo {pages} {011043} (\bibinfo {year}
  {2020})}\BibitemShut {NoStop}%
\bibitem [{\citenamefont {Natsheh}\ \emph {et~al.}(2021)\citenamefont
  {Natsheh}, \citenamefont {Gambassi},\ and\ \citenamefont
  {Mitra}}]{Mitra2021PRB}%
  \BibitemOpen
  \bibfield  {author} {\bibinfo {author} {\bibfnamefont {Muath}\ \bibnamefont
  {Natsheh}}, \bibinfo {author} {\bibfnamefont {Andrea}\ \bibnamefont
  {Gambassi}}, \ and\ \bibinfo {author} {\bibfnamefont {Aditi}\ \bibnamefont
  {Mitra}},\ }\bibfield  {title} {\enquote {\bibinfo {title} {Critical
  properties of the {Floquet} time crystal within the {Gaussian}
  approximation},}\ }\href@noop {} {\bibfield  {journal} {\bibinfo  {journal}
  {Phys. Rev. B}\ }\textbf {\bibinfo {volume} {103}},\ \bibinfo {pages}
  {014305} (\bibinfo {year} {2021})}\BibitemShut {NoStop}%
\bibitem [{\citenamefont {Pizzi}\ \emph {et~al.}(2021)\citenamefont {Pizzi},
  \citenamefont {Knolle},\ and\ \citenamefont {Nunnenkamp}}]{Pizzi2021NP}%
  \BibitemOpen
  \bibfield  {author} {\bibinfo {author} {\bibfnamefont {Andrea}\ \bibnamefont
  {Pizzi}}, \bibinfo {author} {\bibfnamefont {Johannes}\ \bibnamefont
  {Knolle}}, \ and\ \bibinfo {author} {\bibfnamefont {Andreas}\ \bibnamefont
  {Nunnenkamp}},\ }\bibfield  {title} {\enquote {\bibinfo {title} {Higher-order
  and fractional discrete time crystals in clean long-range interacting
  systems},}\ }\href@noop {} {\bibfield  {journal} {\bibinfo  {journal} {Nat.
  Comm.}\ }\textbf {\bibinfo {volume} {12}},\ \bibinfo {pages} {2341} (\bibinfo
  {year} {2021})}\BibitemShut {NoStop}%
\bibitem [{\citenamefont {Sacha}\ and\ \citenamefont
  {Zakrzewski}(2018)}]{Sacha2018RPP}%
  \BibitemOpen
  \bibfield  {author} {\bibinfo {author} {\bibfnamefont {Krzysztof}\
  \bibnamefont {Sacha}}\ and\ \bibinfo {author} {\bibfnamefont {Jakob}\
  \bibnamefont {Zakrzewski}},\ }\bibfield  {title} {\enquote {\bibinfo {title}
  {Time crystals: a review},}\ }\href@noop {} {\bibfield  {journal} {\bibinfo
  {journal} {Rep. Prog. Phys.}\ }\textbf {\bibinfo {volume} {81}},\ \bibinfo
  {pages} {016401} (\bibinfo {year} {2018})}\BibitemShut {NoStop}%
\bibitem [{\citenamefont {Khemani}\ \emph {et~al.}(2019)\citenamefont
  {Khemani}, \citenamefont {Moessner},\ and\ \citenamefont
  {Sondhi}}]{Khemani2019arXiv}%
  \BibitemOpen
  \bibfield  {author} {\bibinfo {author} {\bibfnamefont {Vedika}\ \bibnamefont
  {Khemani}}, \bibinfo {author} {\bibfnamefont {Roderich}\ \bibnamefont
  {Moessner}}, \ and\ \bibinfo {author} {\bibfnamefont {S.~L.}\ \bibnamefont
  {Sondhi}},\ }\href@noop {} {\enquote {\bibinfo {title} {A brief history of
  time crystals},}\ } (\bibinfo {year} {2019}),\ \bibinfo {note} {arXiv:
  1910.10745}\BibitemShut {NoStop}%
\bibitem [{\citenamefont {Else}\ \emph {et~al.}(2020)\citenamefont {Else},
  \citenamefont {Monroe}, \citenamefont {Nayak},\ and\ \citenamefont
  {Yao}}]{Yao2020ARCMP}%
  \BibitemOpen
  \bibfield  {author} {\bibinfo {author} {\bibfnamefont {Dominic~V.}\
  \bibnamefont {Else}}, \bibinfo {author} {\bibfnamefont {Christopher}\
  \bibnamefont {Monroe}}, \bibinfo {author} {\bibfnamefont {Chetan}\
  \bibnamefont {Nayak}}, \ and\ \bibinfo {author} {\bibfnamefont {Norman}\
  \bibnamefont {Yao}},\ }\bibfield  {title} {\enquote {\bibinfo {title}
  {Discrete time crystals},}\ }\href@noop {} {\bibfield  {journal} {\bibinfo
  {journal} {Ann. Rev. Cond. Matt. Phys.}\ }\textbf {\bibinfo {volume} {11}},\
  \bibinfo {pages} {467} (\bibinfo {year} {2020})}\BibitemShut {NoStop}%
\bibitem [{\citenamefont {Sacha}(2020)}]{SachaBook2020}%
  \BibitemOpen
  \bibfield  {author} {\bibinfo {author} {\bibfnamefont {Krzysztof}\
  \bibnamefont {Sacha}},\ }\href@noop {} {\emph {\bibinfo {title} {Time
  Crystals}}}\ (\bibinfo  {publisher} {Springer},\ \bibinfo {year}
  {2020})\BibitemShut {NoStop}%
\bibitem [{\citenamefont {D'Alessio}\ and\ \citenamefont
  {Rigol}(2014)}]{Marcos2014PRX}%
  \BibitemOpen
  \bibfield  {author} {\bibinfo {author} {\bibfnamefont {Luca}\ \bibnamefont
  {D'Alessio}}\ and\ \bibinfo {author} {\bibfnamefont {Marcos}\ \bibnamefont
  {Rigol}},\ }\bibfield  {title} {\enquote {\bibinfo {title} {Long-time
  behavior of isolated periodically driven interacting lattice systems},}\
  }\href@noop {} {\bibfield  {journal} {\bibinfo  {journal} {Phys. Rev. X}\
  }\textbf {\bibinfo {volume} {4}},\ \bibinfo {pages} {041048} (\bibinfo {year}
  {2014})}\BibitemShut {NoStop}%
\bibitem [{\citenamefont {Lazarides}\ \emph {et~al.}(2014)\citenamefont
  {Lazarides}, \citenamefont {Das},\ and\ \citenamefont
  {Moessner}}]{Moessner2014PRE}%
  \BibitemOpen
  \bibfield  {author} {\bibinfo {author} {\bibfnamefont {Achilleas}\
  \bibnamefont {Lazarides}}, \bibinfo {author} {\bibfnamefont {Arnab}\
  \bibnamefont {Das}}, \ and\ \bibinfo {author} {\bibfnamefont {Roderich}\
  \bibnamefont {Moessner}},\ }\bibfield  {title} {\enquote {\bibinfo {title}
  {Equilibrium states of generic quantum systems subject to periodic
  driving},}\ }\href@noop {} {\bibfield  {journal} {\bibinfo  {journal} {Phys.
  Rev. E}\ }\textbf {\bibinfo {volume} {90}},\ \bibinfo {pages} {012110}
  (\bibinfo {year} {2014})}\BibitemShut {NoStop}%
\bibitem [{\citenamefont {Ponte}\ \emph
  {et~al.}(2015{\natexlab{a}})\citenamefont {Ponte}, \citenamefont {Chandrana},
  \citenamefont {Papi{\'c}},\ and\ \citenamefont {Abanin}}]{Abanin2015AnnPhys}%
  \BibitemOpen
  \bibfield  {author} {\bibinfo {author} {\bibfnamefont {Pedro}\ \bibnamefont
  {Ponte}}, \bibinfo {author} {\bibfnamefont {Anushya}\ \bibnamefont
  {Chandrana}}, \bibinfo {author} {\bibfnamefont {Z.}~\bibnamefont
  {Papi{\'c}}}, \ and\ \bibinfo {author} {\bibfnamefont {Dmitry~A.}\
  \bibnamefont {Abanin}},\ }\bibfield  {title} {\enquote {\bibinfo {title}
  {Periodically driven ergodic and many-body localized quantum systems},}\
  }\href@noop {} {\bibfield  {journal} {\bibinfo  {journal} {Ann. Phys.}\
  }\textbf {\bibinfo {volume} {353}},\ \bibinfo {pages} {196} (\bibinfo {year}
  {2015}{\natexlab{a}})}\BibitemShut {NoStop}%
\bibitem [{\citenamefont {Ponte}\ \emph
  {et~al.}(2015{\natexlab{b}})\citenamefont {Ponte}, \citenamefont
  {Papi\ifmmode~\acute{c}\else \'{c}\fi{}}, \citenamefont {Huveneers},\ and\
  \citenamefont {Abanin}}]{Abanin2015PRL}%
  \BibitemOpen
  \bibfield  {author} {\bibinfo {author} {\bibfnamefont {Pedro}\ \bibnamefont
  {Ponte}}, \bibinfo {author} {\bibfnamefont {Z.}~\bibnamefont
  {Papi\ifmmode~\acute{c}\else \'{c}\fi{}}}, \bibinfo {author} {\bibfnamefont
  {Fran\ifmmode \mbox{\c{c}}\else~\c{c}\fi{}ois}\ \bibnamefont {Huveneers}}, \
  and\ \bibinfo {author} {\bibfnamefont {Dmitry~A.}\ \bibnamefont {Abanin}},\
  }\bibfield  {title} {\enquote {\bibinfo {title} {Many-body localization in
  periodically driven systems},}\ }\href@noop {} {\bibfield  {journal}
  {\bibinfo  {journal} {Phys. Rev. Lett.}\ }\textbf {\bibinfo {volume} {114}},\
  \bibinfo {pages} {140401} (\bibinfo {year} {2015}{\natexlab{b}})}\BibitemShut
  {NoStop}%
\bibitem [{\citenamefont {Lazarides}\ \emph {et~al.}(2015)\citenamefont
  {Lazarides}, \citenamefont {Das},\ and\ \citenamefont
  {Moessner}}]{Moessner2015PRL}%
  \BibitemOpen
  \bibfield  {author} {\bibinfo {author} {\bibfnamefont {Achilleas}\
  \bibnamefont {Lazarides}}, \bibinfo {author} {\bibfnamefont {Arnab}\
  \bibnamefont {Das}}, \ and\ \bibinfo {author} {\bibfnamefont {Roderich}\
  \bibnamefont {Moessner}},\ }\bibfield  {title} {\enquote {\bibinfo {title}
  {Fate of many-body localization under periodic driving},}\ }\href@noop {}
  {\bibfield  {journal} {\bibinfo  {journal} {Phys. Rev. Lett.}\ }\textbf
  {\bibinfo {volume} {115}},\ \bibinfo {pages} {030402} (\bibinfo {year}
  {2015})}\BibitemShut {NoStop}%
\bibitem [{\citenamefont {Abanin}\ \emph {et~al.}(2016)\citenamefont {Abanin},
  \citenamefont {Roeck},\ and\ \citenamefont {Huveneers}}]{Abanin2016AnnPhys}%
  \BibitemOpen
  \bibfield  {author} {\bibinfo {author} {\bibfnamefont {Dmitry}\ \bibnamefont
  {Abanin}}, \bibinfo {author} {\bibfnamefont {Wojciech~De}\ \bibnamefont
  {Roeck}}, \ and\ \bibinfo {author} {\bibfnamefont {Fran{\c c}ois}\
  \bibnamefont {Huveneers}},\ }\bibfield  {title} {\enquote {\bibinfo {title}
  {A theory of many-body localization in periodically driven systems},}\
  }\href@noop {} {\bibfield  {journal} {\bibinfo  {journal} {Ann. Phys.}\
  }\textbf {\bibinfo {volume} {372}},\ \bibinfo {pages} {1--11} (\bibinfo
  {year} {2016})}\BibitemShut {NoStop}%
\bibitem [{\citenamefont {Kuwahara}\ \emph {et~al.}(2016)\citenamefont
  {Kuwahara}, \citenamefont {Mori},\ and\ \citenamefont
  {Saito}}]{Kuwahara2015AnnPhys}%
  \BibitemOpen
  \bibfield  {author} {\bibinfo {author} {\bibfnamefont {Tomotaka}\
  \bibnamefont {Kuwahara}}, \bibinfo {author} {\bibfnamefont {Takashi}\
  \bibnamefont {Mori}}, \ and\ \bibinfo {author} {\bibfnamefont {Keiji}\
  \bibnamefont {Saito}},\ }\bibfield  {title} {\enquote {\bibinfo {title}
  {{Floquet-Magnus} theory and generic transient dynamics in periodically
  driven many-body quantum systems},}\ }\href@noop {} {\bibfield  {journal}
  {\bibinfo  {journal} {Ann. Phys.}\ }\textbf {\bibinfo {volume} {367}},\
  \bibinfo {pages} {96} (\bibinfo {year} {2016})}\BibitemShut {NoStop}%
\bibitem [{\citenamefont {Canovi}\ \emph {et~al.}(2016)\citenamefont {Canovi},
  \citenamefont {Kollar},\ and\ \citenamefont {Eckstein}}]{Canovi2016PRE}%
  \BibitemOpen
  \bibfield  {author} {\bibinfo {author} {\bibfnamefont {Elena}\ \bibnamefont
  {Canovi}}, \bibinfo {author} {\bibfnamefont {Marcus}\ \bibnamefont {Kollar}},
  \ and\ \bibinfo {author} {\bibfnamefont {Martin}\ \bibnamefont {Eckstein}},\
  }\bibfield  {title} {\enquote {\bibinfo {title} {Stroboscopic
  prethermalization in weakly interacting periodically driven systems},}\
  }\href@noop {} {\bibfield  {journal} {\bibinfo  {journal} {Phys. Rev. E}\
  }\textbf {\bibinfo {volume} {93}},\ \bibinfo {pages} {012130} (\bibinfo
  {year} {2016})}\BibitemShut {NoStop}%
\bibitem [{\citenamefont {Machado}\ \emph {et~al.}(2019)\citenamefont
  {Machado}, \citenamefont {Kahanamoku-Meyer}, \citenamefont {Else},
  \citenamefont {Nayak},\ and\ \citenamefont {Yao}}]{Yao2019PRR}%
  \BibitemOpen
  \bibfield  {author} {\bibinfo {author} {\bibfnamefont {Francisco}\
  \bibnamefont {Machado}}, \bibinfo {author} {\bibfnamefont {Gregory~D.}\
  \bibnamefont {Kahanamoku-Meyer}}, \bibinfo {author} {\bibfnamefont
  {Dominic~V.}\ \bibnamefont {Else}}, \bibinfo {author} {\bibfnamefont
  {Chetan}\ \bibnamefont {Nayak}}, \ and\ \bibinfo {author} {\bibfnamefont
  {Norman~Y.}\ \bibnamefont {Yao}},\ }\bibfield  {title} {\enquote {\bibinfo
  {title} {Exponentially slow heating in short and long-range interacting
  {Floquet} systems},}\ }\href@noop {} {\bibfield  {journal} {\bibinfo
  {journal} {Phys. Rev. Research}\ }\textbf {\bibinfo {volume} {1}},\ \bibinfo
  {pages} {033202} (\bibinfo {year} {2019})}\BibitemShut {NoStop}%
\bibitem [{\citenamefont {Turner}\ \emph {et~al.}(2018)\citenamefont {Turner},
  \citenamefont {Michailidis}, \citenamefont {Abanin}, \citenamefont {Serbyn},\
  and\ \citenamefont {Papi{\'c}}}]{Papic2018NP}%
  \BibitemOpen
  \bibfield  {author} {\bibinfo {author} {\bibfnamefont {C.~J.}\ \bibnamefont
  {Turner}}, \bibinfo {author} {\bibfnamefont {A.~A.}\ \bibnamefont
  {Michailidis}}, \bibinfo {author} {\bibfnamefont {D.~A.}\ \bibnamefont
  {Abanin}}, \bibinfo {author} {\bibfnamefont {M.}~\bibnamefont {Serbyn}}, \
  and\ \bibinfo {author} {\bibfnamefont {Z.}~\bibnamefont {Papi{\'c}}},\
  }\bibfield  {title} {\enquote {\bibinfo {title} {Weak ergodicity breaking
  from quantum many-body scars},}\ }\href@noop {} {\bibfield  {journal}
  {\bibinfo  {journal} {Nat. Phys.}\ }\textbf {\bibinfo {volume} {14}},\
  \bibinfo {pages} {745} (\bibinfo {year} {2018})}\BibitemShut {NoStop}%
\bibitem [{\citenamefont {Michailidis}\ \emph {et~al.}(2020)\citenamefont
  {Michailidis}, \citenamefont {Turner}, \citenamefont
  {Papi\ifmmode~\acute{c}\else \'{c}\fi{}}, \citenamefont {Abanin},\ and\
  \citenamefont {Serbyn}}]{Serbyn2020PRX}%
  \BibitemOpen
  \bibfield  {author} {\bibinfo {author} {\bibfnamefont {A.~A.}\ \bibnamefont
  {Michailidis}}, \bibinfo {author} {\bibfnamefont {C.~J.}\ \bibnamefont
  {Turner}}, \bibinfo {author} {\bibfnamefont {Z.}~\bibnamefont
  {Papi\ifmmode~\acute{c}\else \'{c}\fi{}}}, \bibinfo {author} {\bibfnamefont
  {D.~A.}\ \bibnamefont {Abanin}}, \ and\ \bibinfo {author} {\bibfnamefont
  {M.}~\bibnamefont {Serbyn}},\ }\bibfield  {title} {\enquote {\bibinfo {title}
  {Slow quantum thermalization and many-body revivals from mixed phase
  space},}\ }\href@noop {} {\bibfield  {journal} {\bibinfo  {journal} {Phys.
  Rev. X}\ }\textbf {\bibinfo {volume} {10}},\ \bibinfo {pages} {011055}
  (\bibinfo {year} {2020})}\BibitemShut {NoStop}%
\bibitem [{\citenamefont {Surace}\ \emph {et~al.}(2021)\citenamefont {Surace},
  \citenamefont {Votto}, \citenamefont {Lazo}, \citenamefont {Silva},
  \citenamefont {Dalmonte},\ and\ \citenamefont {Giudici}}]{GiudiciPRB2021}%
  \BibitemOpen
  \bibfield  {author} {\bibinfo {author} {\bibfnamefont {Federica~Maria}\
  \bibnamefont {Surace}}, \bibinfo {author} {\bibfnamefont {Matteo}\
  \bibnamefont {Votto}}, \bibinfo {author} {\bibfnamefont {Eduardo~Gonzalez}\
  \bibnamefont {Lazo}}, \bibinfo {author} {\bibfnamefont {Alessandro}\
  \bibnamefont {Silva}}, \bibinfo {author} {\bibfnamefont {Marcello}\
  \bibnamefont {Dalmonte}}, \ and\ \bibinfo {author} {\bibfnamefont {Giuliano}\
  \bibnamefont {Giudici}},\ }\bibfield  {title} {\enquote {\bibinfo {title}
  {Exact many-body scars and their stability in constrained quantum chains},}\
  }\href@noop {} {\bibfield  {journal} {\bibinfo  {journal} {Phys. Rev. B}\
  }\textbf {\bibinfo {volume} {103}},\ \bibinfo {pages} {104302} (\bibinfo
  {year} {2021})}\BibitemShut {NoStop}%
\bibitem [{\citenamefont {Pizzi}\ \emph {et~al.}(2020)\citenamefont {Pizzi},
  \citenamefont {Malz}, \citenamefont {De~Tomasi}, \citenamefont {Knolle},\
  and\ \citenamefont {Nunnenkamp}}]{Pizz2020iPRB}%
  \BibitemOpen
  \bibfield  {author} {\bibinfo {author} {\bibfnamefont {Andrea}\ \bibnamefont
  {Pizzi}}, \bibinfo {author} {\bibfnamefont {Daniel}\ \bibnamefont {Malz}},
  \bibinfo {author} {\bibfnamefont {Giuseppe}\ \bibnamefont {De~Tomasi}},
  \bibinfo {author} {\bibfnamefont {Johannes}\ \bibnamefont {Knolle}}, \ and\
  \bibinfo {author} {\bibfnamefont {Andreas}\ \bibnamefont {Nunnenkamp}},\
  }\bibfield  {title} {\enquote {\bibinfo {title} {Time crystallinity and
  finite-size effects in clean {Floquet} systems},}\ }\href@noop {} {\bibfield
  {journal} {\bibinfo  {journal} {Phys. Rev. B}\ }\textbf {\bibinfo {volume}
  {102}},\ \bibinfo {pages} {214207} (\bibinfo {year} {2020})}\BibitemShut
  {NoStop}%
\bibitem [{\citenamefont {Shirley}(1965)}]{ShirleyPR1965}%
  \BibitemOpen
  \bibfield  {author} {\bibinfo {author} {\bibfnamefont {Jon~H.}\ \bibnamefont
  {Shirley}},\ }\bibfield  {title} {\enquote {\bibinfo {title} {Solution of the
  {Schr\"odinger} equation with a {Hamiltonian} periodic in time},}\
  }\href@noop {} {\bibfield  {journal} {\bibinfo  {journal} {Phys. Rev.}\
  }\textbf {\bibinfo {volume} {138}},\ \bibinfo {pages} {B979--B987} (\bibinfo
  {year} {1965})}\BibitemShut {NoStop}%
\bibitem [{\citenamefont {Sambe}(1973)}]{SambePRA1973}%
  \BibitemOpen
  \bibfield  {author} {\bibinfo {author} {\bibfnamefont {Hideo}\ \bibnamefont
  {Sambe}},\ }\bibfield  {title} {\enquote {\bibinfo {title} {Steady states and
  quasienergies of a quantum-mechanical system in an oscillating field},}\
  }\href@noop {} {\bibfield  {journal} {\bibinfo  {journal} {Phys. Rev. A}\
  }\textbf {\bibinfo {volume} {7}},\ \bibinfo {pages} {2203--2213} (\bibinfo
  {year} {1973})}\BibitemShut {NoStop}%
\bibitem [{\citenamefont {Eckardt}\ and\ \citenamefont
  {Anisimovas}(2015)}]{EckardtNJP2015}%
  \BibitemOpen
  \bibfield  {author} {\bibinfo {author} {\bibfnamefont {Andr{\'e}}\
  \bibnamefont {Eckardt}}\ and\ \bibinfo {author} {\bibfnamefont {Egidijus}\
  \bibnamefont {Anisimovas}},\ }\bibfield  {title} {\enquote {\bibinfo {title}
  {High-frequency approximation for periodically driven quantum systems from a
  {Floquet-space} perspective},}\ }\href@noop {} {\bibfield  {journal}
  {\bibinfo  {journal} {New J. Phys.}\ }\textbf {\bibinfo {volume} {17}},\
  \bibinfo {pages} {093039} (\bibinfo {year} {2015})}\BibitemShut {NoStop}%
\bibitem [{\citenamefont {von Keyserlingk}\ \emph {et~al.}(2016)\citenamefont
  {von Keyserlingk}, \citenamefont {Khemani},\ and\ \citenamefont
  {Sondhi}}]{KhemaniPRB2016}%
  \BibitemOpen
  \bibfield  {author} {\bibinfo {author} {\bibfnamefont {C.~W.}\ \bibnamefont
  {von Keyserlingk}}, \bibinfo {author} {\bibfnamefont {Vedika}\ \bibnamefont
  {Khemani}}, \ and\ \bibinfo {author} {\bibfnamefont {S.~L.}\ \bibnamefont
  {Sondhi}},\ }\bibfield  {title} {\enquote {\bibinfo {title} {Absolute
  stability and spatiotemporal long-range order in {Floquet} systems},}\
  }\href@noop {} {\bibfield  {journal} {\bibinfo  {journal} {Phys. Rev. B}\
  }\textbf {\bibinfo {volume} {94}},\ \bibinfo {pages} {085112} (\bibinfo
  {year} {2016})}\BibitemShut {NoStop}%
\bibitem [{\citenamefont {von Keyserlingk}\ and\ \citenamefont
  {Sondhi}(2016)}]{SondhiPRB2016}%
  \BibitemOpen
  \bibfield  {author} {\bibinfo {author} {\bibfnamefont {C.~W.}\ \bibnamefont
  {von Keyserlingk}}\ and\ \bibinfo {author} {\bibfnamefont {S.~L.}\
  \bibnamefont {Sondhi}},\ }\bibfield  {title} {\enquote {\bibinfo {title}
  {Phase structure of one-dimensional interacting {Floquet} systems. ii.
  symmetry-broken phases},}\ }\href@noop {} {\bibfield  {journal} {\bibinfo
  {journal} {Phys. Rev. B}\ }\textbf {\bibinfo {volume} {93}},\ \bibinfo
  {pages} {245146} (\bibinfo {year} {2016})}\BibitemShut {NoStop}%
\bibitem [{\citenamefont {Moessner}\ and\ \citenamefont
  {Sondhi}(2017)}]{MoessnerNP2017}%
  \BibitemOpen
  \bibfield  {author} {\bibinfo {author} {\bibfnamefont {R.}~\bibnamefont
  {Moessner}}\ and\ \bibinfo {author} {\bibfnamefont {S.~L.}\ \bibnamefont
  {Sondhi}},\ }\bibfield  {title} {\enquote {\bibinfo {title} {Equilibration
  and order in quantum {Floquet} matter},}\ }\href@noop {} {\bibfield
  {journal} {\bibinfo  {journal} {Nat. Phys.}\ }\textbf {\bibinfo {volume}
  {13}},\ \bibinfo {pages} {424} (\bibinfo {year} {2017})}\BibitemShut
  {NoStop}%
\bibitem [{\citenamefont {Giergiel}\ \emph {et~al.}(2018)\citenamefont
  {Giergiel}, \citenamefont {Kosior}, \citenamefont {Hannaford},\ and\
  \citenamefont {Sacha}}]{Giergiel2018PRA}%
  \BibitemOpen
  \bibfield  {author} {\bibinfo {author} {\bibfnamefont {Krzysztof}\
  \bibnamefont {Giergiel}}, \bibinfo {author} {\bibfnamefont {Arkadiusz}\
  \bibnamefont {Kosior}}, \bibinfo {author} {\bibfnamefont {Peter}\
  \bibnamefont {Hannaford}}, \ and\ \bibinfo {author} {\bibfnamefont
  {Krzysztof}\ \bibnamefont {Sacha}},\ }\bibfield  {title} {\enquote {\bibinfo
  {title} {Time crystals: Analysis of experimental conditions},}\ }\href@noop
  {} {\bibfield  {journal} {\bibinfo  {journal} {Phys. Rev. A}\ }\textbf
  {\bibinfo {volume} {98}},\ \bibinfo {pages} {013613} (\bibinfo {year}
  {2018})}\BibitemShut {NoStop}%
\bibitem [{\citenamefont {Giergiel}\ \emph {et~al.}(2020)\citenamefont
  {Giergiel}, \citenamefont {Tran}, \citenamefont {Zaheer}, \citenamefont
  {Singh}, \citenamefont {Sidorov}, \citenamefont {Sacha},\ and\ \citenamefont
  {Hannaford}}]{Giergiel2020NJP}%
  \BibitemOpen
  \bibfield  {author} {\bibinfo {author} {\bibfnamefont {Krzysztof}\
  \bibnamefont {Giergiel}}, \bibinfo {author} {\bibfnamefont {Tien}\
  \bibnamefont {Tran}}, \bibinfo {author} {\bibfnamefont {Ali}\ \bibnamefont
  {Zaheer}}, \bibinfo {author} {\bibfnamefont {Arpana}\ \bibnamefont {Singh}},
  \bibinfo {author} {\bibfnamefont {Andrei}\ \bibnamefont {Sidorov}}, \bibinfo
  {author} {\bibfnamefont {Krzysztof}\ \bibnamefont {Sacha}}, \ and\ \bibinfo
  {author} {\bibfnamefont {Peter}\ \bibnamefont {Hannaford}},\ }\bibfield
  {title} {\enquote {\bibinfo {title} {Creating big time crystals with
  ultracold atoms},}\ }\href@noop {} {\bibfield  {journal} {\bibinfo  {journal}
  {New J. Phys.}\ }\textbf {\bibinfo {volume} {22}},\ \bibinfo {pages} {085004}
  (\bibinfo {year} {2020})}\BibitemShut {NoStop}%
\bibitem [{\citenamefont {Kuro{\'s}}\ \emph {et~al.}(2020)\citenamefont
  {Kuro{\'s}}, \citenamefont {Mukherjee}, \citenamefont {Golletz},
  \citenamefont {Sauvage}, \citenamefont {Giergiel}, \citenamefont {Mintert},\
  and\ \citenamefont {Sacha}}]{KurosNJP2020}%
  \BibitemOpen
  \bibfield  {author} {\bibinfo {author} {\bibfnamefont {Arkadiusz}\
  \bibnamefont {Kuro{\'s}}}, \bibinfo {author} {\bibfnamefont {Rick}\
  \bibnamefont {Mukherjee}}, \bibinfo {author} {\bibfnamefont {Weronika}\
  \bibnamefont {Golletz}}, \bibinfo {author} {\bibfnamefont {Frederic}\
  \bibnamefont {Sauvage}}, \bibinfo {author} {\bibfnamefont {Krzysztof}\
  \bibnamefont {Giergiel}}, \bibinfo {author} {\bibfnamefont {Florian}\
  \bibnamefont {Mintert}}, \ and\ \bibinfo {author} {\bibfnamefont {Krzysztof}\
  \bibnamefont {Sacha}},\ }\bibfield  {title} {\enquote {\bibinfo {title}
  {Phase diagram and optimal control for n-tupling discrete time crystal},}\
  }\href@noop {} {\bibfield  {journal} {\bibinfo  {journal} {New J. Phys.}\
  }\textbf {\bibinfo {volume} {22}},\ \bibinfo {pages} {095001} (\bibinfo
  {year} {2020})}\BibitemShut {NoStop}%
\bibitem [{\citenamefont {Wang}\ \emph {et~al.}(2021)\citenamefont {Wang},
  \citenamefont {Hannaford},\ and\ \citenamefont {Dalton}}]{WangNJP2021}%
  \BibitemOpen
  \bibfield  {author} {\bibinfo {author} {\bibfnamefont {Jia}\ \bibnamefont
  {Wang}}, \bibinfo {author} {\bibfnamefont {Peter}\ \bibnamefont {Hannaford}},
  \ and\ \bibinfo {author} {\bibfnamefont {Bryan~J}\ \bibnamefont {Dalton}},\
  }\bibfield  {title} {\enquote {\bibinfo {title} {Many-body effects and
  quantum fluctuations for discrete time crystals in {Bose-Einstein}
  condensates},}\ }\href@noop {} {\bibfield  {journal} {\bibinfo  {journal}
  {New J. Phys.}\ ,\ \bibinfo {pages} {063012}} (\bibinfo {year}
  {2021})}\BibitemShut {NoStop}%
\bibitem [{\citenamefont {Albiez}\ \emph {et~al.}(2005)\citenamefont {Albiez},
  \citenamefont {Gati}, \citenamefont {F\"olling}, \citenamefont {Hunsmann},
  \citenamefont {Cristiani},\ and\ \citenamefont {Oberthaler}}]{Markus2005PRL}%
  \BibitemOpen
  \bibfield  {author} {\bibinfo {author} {\bibfnamefont {Michael}\ \bibnamefont
  {Albiez}}, \bibinfo {author} {\bibfnamefont {Rudolf}\ \bibnamefont {Gati}},
  \bibinfo {author} {\bibfnamefont {Jonas}\ \bibnamefont {F\"olling}}, \bibinfo
  {author} {\bibfnamefont {Stefan}\ \bibnamefont {Hunsmann}}, \bibinfo {author}
  {\bibfnamefont {Matteo}\ \bibnamefont {Cristiani}}, \ and\ \bibinfo {author}
  {\bibfnamefont {Markus~K.}\ \bibnamefont {Oberthaler}},\ }\bibfield  {title}
  {\enquote {\bibinfo {title} {Direct observation of tunneling and nonlinear
  self-trapping in a single bosonic {Josephson} junction},}\ }\href@noop {}
  {\bibfield  {journal} {\bibinfo  {journal} {Phys. Rev. Lett.}\ }\textbf
  {\bibinfo {volume} {95}},\ \bibinfo {pages} {010402} (\bibinfo {year}
  {2005})}\BibitemShut {NoStop}%
\bibitem [{\citenamefont {Buchleitner}\ \emph {et~al.}(2002)\citenamefont
  {Buchleitner}, \citenamefont {Delande},\ and\ \citenamefont
  {Zakrzewski}}]{Zakrzewski2002PR}%
  \BibitemOpen
  \bibfield  {author} {\bibinfo {author} {\bibfnamefont {Andreas}\ \bibnamefont
  {Buchleitner}}, \bibinfo {author} {\bibfnamefont {Dominique}\ \bibnamefont
  {Delande}}, \ and\ \bibinfo {author} {\bibfnamefont {Jakub}\ \bibnamefont
  {Zakrzewski}},\ }\bibfield  {title} {\enquote {\bibinfo {title}
  {Non-dispersive wave packets in periodically driven quantum systems},}\
  }\href@noop {} {\bibfield  {journal} {\bibinfo  {journal} {Phys. Rep.}\
  }\textbf {\bibinfo {volume} {368}},\ \bibinfo {pages} {409} (\bibinfo {year}
  {2002})}\BibitemShut {NoStop}%
\bibitem [{\citenamefont {Lichtenberg}\ and\ \citenamefont
  {Lieberman}(1992)}]{LichtenbergBook1992}%
  \BibitemOpen
  \bibfield  {author} {\bibinfo {author} {\bibfnamefont {A.~J.}\ \bibnamefont
  {Lichtenberg}}\ and\ \bibinfo {author} {\bibfnamefont {M.A.}\ \bibnamefont
  {Lieberman}},\ }\href@noop {} {\emph {\bibinfo {title} {Regular and Chaotic
  Dynamics}}}\ (\bibinfo  {publisher} {Springer},\ \bibinfo {year}
  {1992})\BibitemShut {NoStop}%
\bibitem [{\citenamefont {Dalton}\ \emph {et~al.}(2015)\citenamefont {Dalton},
  \citenamefont {Jeffers},\ and\ \citenamefont {Barnett}}]{DaltonBook2015}%
  \BibitemOpen
  \bibfield  {author} {\bibinfo {author} {\bibfnamefont {B~J}\ \bibnamefont
  {Dalton}}, \bibinfo {author} {\bibfnamefont {J}~\bibnamefont {Jeffers}}, \
  and\ \bibinfo {author} {\bibfnamefont {S~M}\ \bibnamefont {Barnett}},\
  }\href@noop {} {\emph {\bibinfo {title} {Phase Space Methods for Degenerate
  Quantum Gases}}}\ (\bibinfo  {publisher} {Oxford: Oxford University Press},\
  \bibinfo {year} {2015})\BibitemShut {NoStop}%
\bibitem [{\citenamefont {Ribeiro}\ \emph {et~al.}(2008)\citenamefont
  {Ribeiro}, \citenamefont {Vidal},\ and\ \citenamefont
  {Mosseri}}]{Ribeiro2008PRE}%
  \BibitemOpen
  \bibfield  {author} {\bibinfo {author} {\bibfnamefont {Pedro}\ \bibnamefont
  {Ribeiro}}, \bibinfo {author} {\bibfnamefont {Julien}\ \bibnamefont {Vidal}},
  \ and\ \bibinfo {author} {\bibfnamefont {R\'emy}\ \bibnamefont {Mosseri}},\
  }\bibfield  {title} {\enquote {\bibinfo {title} {Exact spectrum of the
  {Lipkin-Meshkov-Glick} model in the thermodynamic limit and finite-size
  corrections},}\ }\href@noop {} {\bibfield  {journal} {\bibinfo  {journal}
  {Phys. Rev. E}\ }\textbf {\bibinfo {volume} {78}},\ \bibinfo {pages} {021106}
  (\bibinfo {year} {2008})}\BibitemShut {NoStop}%
\bibitem [{\citenamefont {Mazza}\ and\ \citenamefont
  {Fabrizio}(2012)}]{Fabrizio2012PRB}%
  \BibitemOpen
  \bibfield  {author} {\bibinfo {author} {\bibfnamefont {Giacomo}\ \bibnamefont
  {Mazza}}\ and\ \bibinfo {author} {\bibfnamefont {Michele}\ \bibnamefont
  {Fabrizio}},\ }\bibfield  {title} {\enquote {\bibinfo {title} {Dynamical
  quantum phase transitions and broken-symmetry edges in the many-body
  eigenvalue spectrum},}\ }\href@noop {} {\bibfield  {journal} {\bibinfo
  {journal} {Phys. Rev. B}\ }\textbf {\bibinfo {volume} {86}},\ \bibinfo
  {pages} {184303} (\bibinfo {year} {2012})}\BibitemShut {NoStop}%
\bibitem [{\citenamefont {Dalton}\ and\ \citenamefont
  {Ghanbari}(2012)}]{Dalton2012JMO}%
  \BibitemOpen
  \bibfield  {author} {\bibinfo {author} {\bibfnamefont {B.J.}\ \bibnamefont
  {Dalton}}\ and\ \bibinfo {author} {\bibfnamefont {S.}~\bibnamefont
  {Ghanbari}},\ }\bibfield  {title} {\enquote {\bibinfo {title} {Two mode
  theory of {Bose-Einstein} condensates: interferometry and the {Josephson}
  model},}\ }\href@noop {} {\bibfield  {journal} {\bibinfo  {journal} {J. Mod.
  Opt.}\ }\textbf {\bibinfo {volume} {59}},\ \bibinfo {pages} {287} (\bibinfo
  {year} {2012})}\BibitemShut {NoStop}%
\end{thebibliography}%

\end{document}